\documentclass[11pt,english,longbibliography]{article}
\usepackage{url}

\usepackage{draft} 
\usepackage{putex}
\usepackage{comment}

\usepackage{hyperref}

\newcommand{\CO}{{\cal O}}

\newcommand{\CD}{{\cal D}}

\newcommand{\CK}{{\cal K}}

\newcommand{\CF}{{\cal F}}
\renewcommand{\CP}{{\cal P}}

\newcommand{\CI}{{\cal I}}

\newcommand{\ts}{{\widetilde s}}

\newcommand{\CM}{{\cal M}}
\usepackage{amsthm}

\makeatletter
\newcommand*{\rom}[1]{\expandafter\@slowromancap\romannumeral #1@}
\makeatother

\usepackage[latin9]{inputenc}
\usepackage{multirow}
\usepackage{verbatim}
\usepackage{prettyref}
\usepackage[numbers,sort&compress]{natbib}
\usepackage{amsmath}
\usepackage{amssymb}
\usepackage{tikz-cd}
\tikzset{commutative diagrams/row sep/huge=4cm}
\tikzset{commutative diagrams/column sep/huge=4cm}
\tikzcdset{scale cd/.style={every label/.append style={scale=#1},
    cells={nodes={scale=#1}}}}
\usepackage{lmodern}
\usepackage{tcolorbox}
\usepackage{cases}
\usepackage{bbold}
\usepackage{bm}
\usepackage{ytableau}
\usepackage{youngtab}
\usepackage{mathtools}
\usepackage{graphicx}
\usepackage[nottoc]{tocbibind}
\usepackage{mathrsfs}
\usepackage{caption}
\usepackage{subcaption}
\usepackage{cancel}
\usepackage{float}
\usepackage{tikz}

\usetikzlibrary{arrows.meta}
\usetikzlibrary{bending}
\usepackage{tikz}

\tikzset{
    Witten diagram/.style={
        execute at begin picture={
            \draw[blue, line width=1.5pt] circle[radius=\pgfkeysvalueof{/tikz/Witten/radius}];
            \path node (X){\phantom{X}};
        },
        baseline={(X.base)}
    },
    vertex/.style={circle,fill,inner sep=1.5pt,node contents={}},
    Witten/.cd,
    radius/.initial=3cm
}
\usetikzlibrary{patterns}
\usepackage{lipsum}
\usepackage{framed}
\usepackage{adjustbox}
\usepackage{mathtools}
\usepackage{braket}

 \usepackage{slashed}
\usepackage{esvect}
\usepackage{dsfont}

\usepackage{color}
\definecolor{darkgreen}{rgb}{0,0.5,0}
\definecolor{darkblue}{rgb}{0,0,0.6}
\definecolor{purple}{rgb}{0.4,.2,0.7}

\numberwithin{equation}{section}
\numberwithin{figure}{section}
\numberwithin{table}{section}

\def\CG{{\cal G}}

\def\CM{{\cal M}}
\def\CN{{\cal N}}
\def\CD{{\cal D}}
\def\tr{\,{\rm tr}\,}

\DeclareFontShape{OT1}{cmr}{mx}{n}{<->cmr10}{}

\begin{document}


\title{\centering Higher loops in AdS: applications to boundary CFT
}

\authors{Simone Giombi \worksat{\PUJ} and Zimo Sun \worksat{\PUJ,\PUG} }

\institution{PUJ}{Joseph Henry Laboratories, Princeton University, Princeton, NJ 08544, USA}

\institution{PUG}{Princeton Gravity Initiative, Princeton University, Princeton, NJ 08544, USA}

\abstract{The Euclidean Anti-de Sitter (AdS) space provides a natural framework for studying boundary conformal field theory (BCFT).
We analyze the conformal boundary conditions of the critical O$(N)$ model in $d=4-\epsilon$ dimensions using the $\epsilon$-expansion, and extract some BCFT observables through higher-loop calculations in AdS. Specifically, in the so-called ``ordinary" universality class, we determine the free energy to  four-loop order and the one-point function of the lightest O$(N)$ singlet operator to three-loop order. In the  symmetry breaking ``normal" universality class, we derive the two-loop free energy and compute the leading correction to the one-point function of the lightest O($N$) vector. 
We apply Pad\'e approximants to extract the corresponding conformal data in three dimensions. In particular, from a suitable dimensional continuation of the free energy in AdS, we obtain estimates for the boundary central charge of the BCFT in $d=3$.} 

\date{}

\maketitle

\tableofcontents

\section{Introduction}
By virtue of a Weyl transformation, a conformal field theory in flat $d$-dimensional space with a boundary may be mapped to hyperbolic space, or Euclidean AdS.\footnote{We will assume Euclidean signature throughout this paper.} This provides a useful approach to the problem of boundary conformal field theory (BCFT) \cite{Paulos:2016fap, Carmi:2018qzm, Giombi:2020rmc} (see also e.g. \cite{Giombi:2021uae, Giombi:2021cnr, Cuomo:2021kfm, Giombi:2024qbm} for related work on the AdS approach to boundary and defect CFT, and \cite{Ankur:2023lum, Copetti:2023sya, Ciccone:2024guw, Bason:2025zpy} for other recent applications of QFT in AdS). The SO$(d,1)$ residual symmetry of the BCFT is manifest in the AdS approach, and correlation functions take a simple form due to the maximal symmetry of the space. For example, one-point functions of scalar operators are constant in AdS, while two-point functions are constrained to be functions of the geodesic distance. In this paper, continuing and extending the work in \cite{Giombi:2020rmc}, we focus on the O$(N)$ model in the presence of a boundary, and obtain some of its BCFT data by performing higher-loop calculations in AdS in the framework of the $d=4-\epsilon$ expansion.  

In addition to its bulk and boundary local correlation functions, an important quantity in a BCFT in $d=3$ is its boundary central charge $c$. This is related to the boundary conformal anomaly via \cite{Jensen:2015swa, Herzog:2017kkj}
\begin{equation} \label{TraceAnomaly3d}
\langle {T^{\mu}}_{\mu} \rangle^{d = 3} = \frac{\delta(x_{\perp})}{24 \pi} \left( c \hat{{\cal R}} + d_1 \tr \hat{K}^2 \right)
\end{equation}
where $\hat{{\cal R}}$ and $\hat{K}$ are the scalar and extrinsic curvature of the boundary, and $x_{\perp}$ the coordinate perpendicular to the boundary. The choice of normalization above is such that $c=-1/16$ for a free scalar with Dirichlet boundary condition (and $c=+1/16$ for Neumann boundary condition). It was proved in \cite{Jensen:2015swa} that the boundary central charge $c$ satisfies a $c$-theorem $c_{\rm UV}>c_{\rm IR}$ under boundary RG flows (the same result applies more generally to surface defects).\footnote{The coefficient $d_1$ in (\ref{TraceAnomaly3d}) is related to $C_D$ \cite{Herzog:2017kkj}, the normalization of the displacement operator two-point function, and it does not in general satisfy a monotonicity constraint under boundary or surface defect RG flows.} The central charge $c$ may be also extracted from the logarithmic divergent part of the free energy on a round hemisphere with radius $R$, as $F_{HS^3} = -c/3 \log(R\mu)+\ldots$, where $\mu$ is a UV cutoff. Equivalently, it may be obtained from the free energy on the hyperbolic space $H^3$ (which is conformally related to the hemisphere), where the logarithmic divergence now comes from the regularized volume of $H^3$. Within the framework of dimensional regularization, such logarithmic divergence arises as a pole near $d=3$, and the central charge may be extracted from the dimensionally continued quantity \cite{Giombi:2020rmc}
\begin{equation}
\label{tilde-F}
\tilde{F}_{H^d} = -\sin\left(\frac{\pi(d-1)}{2}\right)F_{H^d}\,.
\end{equation} 
Due to the sine factor, this quantity is finite in $d=3$ and equal to $\tilde{F}_{H^d}|_{d=3} = \frac{\pi}{6}c$. It is natural to compute (\ref{tilde-F}) perturbatively in $d=4-\epsilon$ at the Wilson-Fisher fixed point, and use it to estimate the central charge in $d=3$. However, one drawback of (\ref{tilde-F}) is that near even $d$ it contains poles in dimensional regularization, due to the bulk conformal anomaly, which may make its dimensional extrapolation to $d=3$ challenging. In this paper, we will consider instead the following quantity
\begin{equation}
\label{tilde-s}
\tilde{s} =  -\sin\left(\frac{\pi(d-1)}{2}\right)\left(F_{H^d}-\frac{1}{2}F_{S^d}\right)
\end{equation}
where $F_{S^d}$ is the free energy of the CFT on a round sphere. This quantity was previously defined in \cite{Kobayashi:2018lil}, which proposed generalized $c$-theorems in continuous dimension $d$ in defect and boundary CFT. Due to the subtraction of the sphere free energy term, $\tilde{s}$ is free of poles near even $d$, while its limit in $d=3$ (or other odd $d$) still captures the central charge as $\tilde{s}_{d=3}=\frac{\pi}{6}c$ (this is because the sphere free energy $F_{S^d}$ is finite, within dimensional regularization, for all odd $d$). Note also that the $d=2$ limit of $\tilde{s}$ is finite and related to the so-called $g$-function of the BCFT \cite{PhysRevLett.67.161}
\begin{equation}
\label{2d-tilde-s}
\tilde{s}_{d=2} = -F_{H^{2}} + \frac{1}{2} F_{S^{2}} = \log(g)\,.
\end{equation}
Indeed, the $g$-function of a 2$d$ BCFT may be obtained from the partition function on a round hemisphere via $g=Z_{HS^2}/\sqrt{Z_{S^2}}$, which leads to (\ref{2d-tilde-s}) upon identifying $F_{HS^2}$ with $F_{H^2}$ by a conformal transformation.\footnote{In more detail, recall that the $g$-function associated with a conformal invariant boundary state $|\alpha\rangle$ is given by $\langle \Omega|\alpha\rangle/\sqrt{\langle\Omega|\Omega\rangle}$ where $|\Omega\rangle $ denotes the vacuum state \cite{PhysRevLett.67.161, Dorey:1999cj, Dorey:2009vg}. We  identify $\langle\Omega|\Omega\rangle$ with the sphere partition function $Z_{S^2}$, and  $\langle \Omega|\alpha\rangle$ with the hemishphere (or AdS) partition function $Z^\alpha_{HS^2}$ corresponding to the  boundary condition $\alpha$. This yields $\log g_\alpha =\log(Z^\alpha_{HS^2}/\sqrt{Z_{S^2}}) =-F^\alpha_{HS^2}+\frac{1}{2}F_{S^2}=-F^\alpha_{H^2}+\frac{1}{2}F_{S^2}=\tilde s_\alpha$.} In summary, the $\tilde{s}$ function is expected to be a finite quantity for all $d$, interpolating between the boundary central charge in $d=3$ and the $g$-function in $d=2$ (as well as their higher dimensional analogs in higher odd and even $d$).

At a practical level, the $\ts$ function provides a way to estimate the boundary central charge for strongly coupled CFTs in three dimensions using the $\epsilon$-expansion. As mentioned above, in this paper we focus on the critical O$(N)$ model at its Wilson-Fisher fixed point, which is weakly coupled in $d=4-\epsilon$ dimensions. It is well-known that, with the bulk coupling tuned to its critical point, the O$(N)$ model admits various conformal boundary conditions leading to distinct BCFTs connected by boundary RG flows. The possible boundary conditions are easy to understand in the framework of the weakly coupled $\phi^4$ theory in $d=4-\epsilon$. If one requires unbroken O$(N)$ symmetry, one may impose Dirichlet or Neumann boundary conditions on the fundamental field $\phi^I$. These lead to two O$(N)$ preserving BCFTs, known respectively as the ``ordinary" and ``special" universality classes (we refer the reader to \cite{Die97} for a comprehensive review on the O$(N)$ model with a boundary). One may flow from the special to the ordinary BCFTs by adding a boundary mass term. Since the use of Neumann boundary conditions presents certain additional technical difficulties \footnote{With Neumann boundary condition, the loop integrals exhibit additional IR divergences that require special treatment. Up to order $\epsilon^2$, this technical difficulty can be overcome by introducing an IR regular $\delta$ and take the $\delta\to 0 $ limit at the end \cite{Giombi:2020rmc}, as every integral to that order can be evaluated analytically for any $\delta$ and $\epsilon$. To order $\epsilon^3$, however, the IR regulator appears to make the evaluation of  diagrams $(f)$ and $(g)$ in fig. \ref{Orddiagram1} intractable analytically.}, in this paper we will focus on the ``ordinary" (Dirichlet) boundary condition, and compute $\ts$ to order $\epsilon^3$, extending the previous result obtained in \cite{Giombi:2020rmc}. We will also compute the bulk one-point function of the leading operator $\phi^2=\phi^I\phi^I$ to order $\epsilon^2$. As we explain below, the proper definition of the primary $\phi^2$ operator in AdS requires carefully taking into account mixing with the scalar curvature, which is essential in order to get agreement with previous results obtained by analytic bootstrap methods \cite{Bissi:2018mcq,Dey:2016mcs}. The ordinary universality class exists in the whole range $2\le d\le 4$, and in particular for $N=1$ and $d=2$ we can make contact with known data for the $Z_2$ preserving boundary condition in the Ising BCFT, which can be used as a constraint for ``two-sided" Pad\'e approximants. 

In addition to the symmetry preserving boundary conditions, one can also consider breaking of the O$(N)$ symmetry. The simplest option is explicit breaking, obtained by adding a linear perturbation $\sim \phi^{I=N}$ at the boundary (a boundary magnetic field pointing in a chosen direction). This leads to a non-zero bulk one-point function of $\phi^N$, which breaks the O$(N)$ symmetry to O$(N-1)$. The corresponding BCFT is known as the ``normal" universality class. Alternatively, one may consider spontaneous breaking of the O$(N)$ symmetry, which can be obtained by adding a negative mass term at the boundary. This gives rise to the so-called ``extraordinary" class. It is expected that in $3<d<4$ and $N\ge 2$ the normal and extraordinary theories are essentially equivalent \cite{Bray:1977fvl, Burkhardt_1987, PhysRevB.50.3894, https://doi.org/10.1002/bbpc.19940980344} (up to a decoupled sector of boundary Goldstone bosons), while for $N=1$ they are equivalent in $3\le d<4$. The normal boundary condition exists for all $N$ in the whole range $2\le d<4$, while the extraordinary one (as well as the special) do not extend below $d=3$. It was recently argued in \cite{Metlitski:2020cqy} that in $d=3$ and $2\le N <N_c$ (for some critical $N_c$), a modified form of the extraordinary class, dubbed ``extraodinary-log", also exists. In this paper, since we work within the $d=4-\epsilon$ expansion, we will only consider the normal boundary condition. In the AdS approach, it can be identified with the expansion around the minimum of the potential which breaks the O$(N)$ symmetry \cite{Giombi:2020rmc}. We will use this approach to compute $\tilde{s}$ to order $\epsilon^2$, as well as the one-point function of $\phi^N$ to next-to-leading order, extending the previous results obtained in \cite{Giombi:2020rmc}.

For  the $N=1$ case, we apply two-sided Pad\'e resummation to extract the boundary conformal data of the 3$d$ Ising model. At the normal fixed point, the Pad\'e estimates of the boundary central charge $c_{\rm Ising}^{\rm nor}\approx -1.43$ and the one-point coefficient $a_\phi^{\rm nor} \approx 2.71$ nicely agree with the fuzzy sphere results  \cite{Zhou:2024dbt}, which report $c_{\rm Ising}^{\rm nor}= -1.44(6)$ and  $a_\phi^{\rm nor} = 2.58(16)$.
At the ordinary fixed point, the one-point coefficient $a_{\phi^2}^{\rm ord} \approx -0.728$, obtained from the $\epsilon$-expansion, is consistent with the fuzzy sphere result $-0.74(4)$ \cite{Zhou:2024dbt}. However,  the Pad\'e result for the boundary central charge at the ordinary fixed point shows some  tension with the fuzzy sphere data, though it is in good agreement with an alternative $\epsilon$-expansion result computed in the surface defect framework \cite{Diatlyk:2024ngd}.

The rest of this paper is organized as follows: in section \ref{review}, we review the propagator and one-loop free energy of a free scalar field in AdS. We also derive the $\widetilde s$ function for a conformally coupled scalar with the Dirichlet or Neumann boundary condition. In section \ref{ordsec}, we compute the free energy of the critical O$(N)$ model  and the one-point function of the quadratic operator $\phi^2$ with the ordinary boundary condition, and discuss their Pad\'e approximants. In section \ref{extsec}, we analyze the  symmetry breaking boundary condition, computing both the free energy and the one-point function of $\phi^I$. Finally, the appendices contain various technical results, some of which might be of independent interest.

\section{Review of AdS propagators and one-loop free energy}\label{review}
Our strategy in this paper is to map all BCFT calculations onto AdS space. To set up  the calculations in later sections, we begin by reviewing some fundamental aspects of AdS, including its geometry, the propagators, and the one-loop free energy.

The $d$ dimensional Euclidean AdS space is a hypersurface in the ambient space $\mathbb R^{1,d}$:
\begin{align}
-X_0^2+X_1^2+\cdots + X_{d}^2 = -1~,
\end{align}
where the AdS radius is taken to be 1. We will mainly work with 
the coordinates $X^0 = \cosh r, X^a = \sinh r\, \omega^a$, where  $\omega^a\in S^{d-1}$.  The corresponding metric $ds^2 = dr^2+\sinh^2 r\, d\omega^2$ represents a hyperbolic ball.
The AdS volume in dimensional regularization is thus 
\begin{align}
{\rm Vol}(H^{d}) = {\rm Vol}(S^{d-1}) \int_0^\infty dr\sinh^{d-1} r = \pi^{\frac{d-1}{2}}\Gamma\left(\frac{1-d}{2}\right)~.
\end{align}

For a free scalar field of boundary scaling dimension $\Delta$ in AdS$_{d}$, the propagator is a function of $\sigma\equiv -X_1\cdot X_2\in [1,\infty)$
\begin{align}\label{Gd}
G_\Delta(X_1, X_2) \equiv G_\Delta(\sigma)=  \frac{c_\Delta}{\sigma^\Delta} \mathbf{F} \left(\frac{\Delta}{2}, \frac{\Delta+1}{2}, \Delta+\frac{3-d}{2}, \frac{1}{\sigma^2} \right),\quad c_\Delta = \frac{\Gamma(\Delta)}{2^{\Delta+1}\pi^{\frac{d-1}{2}}}~,
\end{align}
where $\mathbf F(a, b, c,z) = 1/\Gamma(c)F(a, b, c,z)$ is the regularized hypergeometric function. 
The coincident limit corresponds to $\sigma=1$. In dimensional regularization,  the coincident limit is given by  
\begin{align}\label{GdXX}
G_\Delta(1) = \frac{\Gamma(1-\frac{d}{2})\Gamma(\Delta)}{(4\pi)^{\frac{d}{2}}\Gamma(2-d+\Delta)} ~.
\end{align}

The one-loop free energy of a scalar field in AdS$_{d}$ is given by 
\begin{align}\label{F1def}
F_1(\Delta) & = \frac{1}{2}\log {\rm Tr}\left(-\nabla^2+\Delta(\Delta-d+1)\right)\nonumber\\
&= \frac{{\rm Vol}(H^{d})}{2(4\pi)^{\frac{d}{2}} \Gamma(\frac{d}{2})}\int_\mathbb{R}d\lambda\frac{\Gamma(\frac{d-1}{2}+ i\lambda)\Gamma(\frac{d-1}{2}- i\lambda)}{\Gamma(i\lambda)\Gamma(-i\lambda)} \log\left(\lambda^2+\left(\Delta-\frac{d-1}{2}\right)^2\right).
\end{align}
This follows from the fact that $-\nabla^2$ has eigenvalues $\lambda^2+\frac{(d-1)^2}{4}$ with spectral density \cite{doi:10.1063/1.530850, Bytsenko:1994bc}
\begin{align}
\mu(\lambda) = \frac{{\rm Vol}(H^{d})}{(4\pi)^{\frac{d}{2}}}\frac{\Gamma(\frac{d-1}{2}+ i\lambda)\Gamma(\frac{d-1}{2}- i\lambda)}{\Gamma(i\lambda)\Gamma(-i\lambda)} ~.
\end{align}
This integral representation of $F_1(\Delta) $ does not admit straightforward analytical evaluation for arbitrary $d$. It is also very hard to obtain  the $\epsilon$-expansion of $F_1(\Delta)$ around $d=4-\epsilon$ using \eqref{F1def} due to the noncompact integration domain \footnote{We will provide a method in appendix \ref{FDexp}.}. On the other hand, it has been found using \eqref{F1def} that $\partial_\Delta F_1(\Delta)$ takes a much simpler form \cite{Giombi:2020rmc}
\begin{align}
\partial_\Delta F_1(\Delta) = \frac{{\rm Vol}(H^{d})}{2(4\pi)^{\frac{d}{2}}}\frac{(2\Delta-d+1)\Gamma(\Delta)\Gamma(1-\frac{d}{2})}{\Gamma(2-d+\Delta)}
\end{align}
which leads to 
\begin{align}\label{F1mD}
 F_1(\Delta) - F_D =  \frac{{\rm Vol}(H^{d})}{2(4\pi)^{\frac{d}{2}}}\int^\Delta_{\frac{d}{2}}\,d\hat\Delta\frac{(2\hat\Delta-d+1)\Gamma(\hat\Delta)\Gamma(1-\frac{d}{2})}{\Gamma(2-d+\hat\Delta)}~,
\end{align}
where $F_D$ denotes the one-loop free energy of a conformally coupled scalar with the Dirichlet boundary condition, i.e. $F_D= F_1(\frac{d}{2})$. We also define $F_N= F_1(\frac{d-2}{2})$,  corresponding to the Neumann boundary condition.  

A convenient way to compute $F_D$ and $F_N$, following the argument of \cite{Giombi:2024qbm}, is as follows. Due to the conformal symmetry, $F_D$($F_N$) is also equal to the hemisphere free energy with the Dirichlet (Neumann) boundary condition. The sum of $F_D$ and $F_N$ thus gives the free energy $F_{\rm conf}$ of a conformally coupled scalar on $S^d$:
\begin{align}\label{Fs}
F_D + F_N=F_{\rm conf}  = -\int_0^1\,du u\sin(\pi u) \frac{\Gamma(\frac{d}{2}+u)\Gamma(\frac{d}{2}-u)}{\Gamma(d+1)\sin\frac{d\pi}{2}}~.
\end{align}
where the integral representation of $F_{\rm conf}$ is given in \cite{Giombi:2014xxa}. On the other hand, the difference between $F_D$ and $F_N$ can be formally viewed as arising from a ``double-trace deformation" (more precisely, it arises from a boundary mass term, but mathematically this is equivalent to a double-trace deformation)
and admits the following integral representation \cite{Giombi:2020rmc}
\begin{align}\label{FDmN}
F_D - F_N= \int_0^\frac{1}{2}\,du u\sin(\pi u) \frac{\Gamma(\frac{d-1}{2}+u)\Gamma(\frac{d-1}{2}-u)}{\Gamma(d)\sin\frac{(d-1)\pi}{2}}~.
\end{align}
Combing \eqref{Fs} and \eqref{FDmN} gives
\begin{align}\label{FDint}
F_D = \int_0^\frac{1}{2}\,du u\sin(\pi u) \frac{\Gamma(\frac{d-1}{2}+u)\Gamma(\frac{d-1}{2}-u)}{2\Gamma(d)\sin\frac{(d-1)\pi}{2}}-\int_0^1\,du u\sin(\pi u) \frac{\Gamma(\frac{d}{2}+u)\Gamma(\frac{d}{2}-u)}{2\Gamma(d+1)\sin\frac{d\pi}{2}}~,
\end{align}
which provides an efficient way to extract the $\epsilon$-expansion of $F_D$, e.g.
\begin{align}\label{FDexp1}
F_D &\stackrel{d=4-\epsilon}{=} \frac{1}{180\epsilon} +\frac{240\log(A)-480\zeta'(-3)-29-16\gamma_E}{2880} - \frac{\zeta(3)}{(4\pi)^2} \nonumber\\
&\qquad - 0.003149\epsilon-0.011698\epsilon^2-0.012781\epsilon^3+\CO(\epsilon^4)~,
\end{align}
where $A$ is the Glaisher's constant and $\gamma_E$ is the Euler-Mascheroni constant. In appendix \ref{FDexp} we check that this agrees with the explicit evaluation of (\ref{F1def}) in AdS$_d$.

\begin{figure}[t]
\centering
\includegraphics[width=14cm]{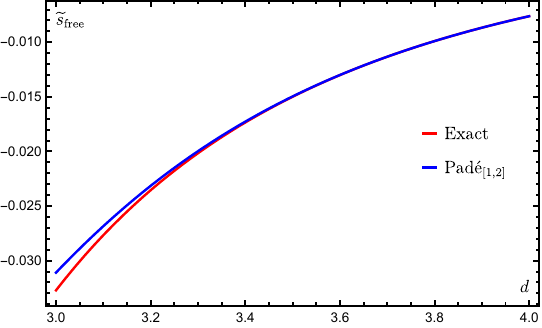}
\caption{The $\widetilde s$ function of a conformally coupled scalar with Dirichlet boundary condition. The red line corresponds to \eqref{tsfree} and the blue line corresponds to the [1,2] Pad\'e approximant of \eqref{sfreeexp}. }
\label{sfree}
\end{figure}

Using \eqref{FDmN}, we can easily write down the $\widetilde s$ function of  a conformally coupled scalar with the Dirichlet boundary condition
\begin{align}\label{tsfree}
\widetilde s_{\rm free} (d)&=  -\sin\left(\frac{d-1}{2}\pi\right)\left(F_D - \frac{1}{2} F_{\rm conf}\right)=-\frac{1}{2}\sin\left(\frac{d-1}{2}\pi\right)(F_D-F_N)\nonumber\\
&=-\frac{1}{2\Gamma(d)}\int_0^\frac{1}{2}\,du u\sin(\pi u) \Gamma\left(\frac{d-1}{2}+u\right)\Gamma\left(\frac{d-1}{2}-u\right)~.
\end{align}
 We plot it as a function of $d\in [3,4]$ in fig. \ref{sfree} (the red line). In $d=3$, $\tilde{s}_{\rm free}$ is equal to $-\frac{\pi}{96}\approx -0.0327$.  The corresponding boundary central charge in the normalization defined by (\ref{TraceAnomaly3d}) is $c_{\rm free} = \frac{6}{\pi}s_{\rm free} (3) = -\frac{1}{16}$, in agreement with \cite{Jensen:2015swa}. For the Neumann boundary condition, we get $c_{\rm free}=\frac{1}{16}$ instead.

On the other hand, to probe the effectiveness of $\tilde{s}$ in the $\epsilon$-expansion, it is interesting to expand $\widetilde s_{\rm free} $ around $d=4$ dimensions and use Pad\'e resummation to estimate its numerical value  in $3d$. For example, keeping terms up to order $\epsilon^3$, we have 
\begin{align}\label{sfreeexp}
\widetilde s_{\rm free} (4-\epsilon)&=  -0.00761211-0.0097764 \epsilon-0.00720205\epsilon^2-0.0041371 \epsilon^3+\CO(\epsilon^4)~.
\end{align}
Its [1,2] Pad\'e approximant is regular for $d\in[3,4]$ as shown in fig. (\ref{sfree}) (the blue line) and yields -0.0311 in 3$d$, which is about 5\% off the exact value. With only four terms in the $\epsilon$-expansion, we get a very good approximation of $\widetilde s_{\rm free}(d)$ for $3\le d\le 4$. 

\section{The ordinary boundary condition}\label{ordsec}
In this section, we compute the $\epsilon$-expansion of the AdS free energy with the Dirichlet boundary condition for the critical O$(N)$ model, generalizing  the results of \cite{Giombi:2020rmc}. Based on the free energy calculation, we also derive the one-point function of the quadratic O$(N)$ invariant operator $\phi^2$.
We then apply Pad\'e resummation to extract the boundary central charge and bulk one-point coefficient in the ordinary universality class of the $3d$ O$(N)$ model.

Mapping the critical  O$(N)$ model to  AdS$_{d}$ gives the following action 
\begin{align}\label{Sor}
S = \int d^{d} x\sqrt{g} \left[\frac{1}{2}\left(\partial\phi^I\right)^2 - \frac{d(d-2)}{8}\phi^I\phi^I+ \frac{\lambda_0}{4}\left(\phi^I\phi^I\right)^2\right]~,
\end{align}
where $I=1,2,\cdots,N$ is an O($N$) index, and $\lambda_0$ the bare coupling constant.
The ordinary phase corresponds to  choosing the Dirichlet boundary condition for the conformally coupled scalars $\phi^I$ in free theory, i.e. 
\begin{align}
\left\langle\phi^I(X)\phi^J(Y)\right\rangle_0 = \delta^{IJ} G_{\frac{d}{2}} (X, Y) = \frac{ \delta^{IJ} \Gamma(\frac{d}{2}-1)}{2^{\frac{d+2}{2}}\pi^{\frac{d}{2}}}\left(\frac{1}{(\sigma-1)^{\frac{d}{2}-1}}- \frac{1}{(\sigma+1)^{\frac{d}{2}-1}}\right)
\end{align}
where $\sigma\equiv - X\cdot Y$. 

\subsection{The three-loop free energy}
In \cite{Giombi:2020rmc}, the AdS free energy of the O$(N)$ model was computed  to the second order in the coupling constant
\begin{align}
F_2^{\rm ord}=N F_D+\frac{\lambda_0}{4}\int_X \left\langle\left(\phi^I\phi^I(X)\right)^2\right\rangle_0 - \frac{\lambda_0^2}{32} \int_X\int_Y\left\langle\left(\phi^I\phi^I(X)\right)^2 \left(\phi^J\phi^J(Y)\right)^2\right\rangle_0+\CO(\lambda_0^3)~, 
\end{align}
where  $\int_X$ is a shorthand notation for $\int d^{d} x\sqrt{g} $ on AdS$_{d}$. In this subsection, we reproduce their results using slightly different methods. We will also keep one more order in the $\epsilon$-expansion compared to \cite{Giombi:2020rmc} since we will  compute the $\lambda_0^3$ terms later.

\begin{figure}[t]
\centering
$(a)$\,\, \begin{tikzpicture} [baseline={(0,0)}]
\draw[color=black] (0,0) circle [radius=0.5]; 
\draw[color=black] (1,0) circle [radius=0.5]; 
\draw (0.5,0) node[vertex]{} ;
\end{tikzpicture} 
\qquad
$(b)$\,\,\begin{tikzpicture} [baseline={(0,0)}]
\draw[color=black] (0,0) circle [radius=0.5]; 
\draw[color=black] (1,0) circle [radius=0.5]; 
\draw[color=black] (2,0) circle [radius=0.5]; 
\draw (0.5,0) node[vertex]{} ;
\draw (1.5,0) node[vertex]{} ;
\end{tikzpicture} 
\qquad
$(c)$\,\,\begin{tikzpicture} [baseline={(0,0)}]
\draw (0,0) node[vertex]{} to [out=-20, in=-160]  (2,0) node[vertex]{} to [out=120, in=60] (0,0) to [out=-60, in=-120] (2,0) to [out=160, in=20] (0,0);
\end{tikzpicture} 
\caption{The two-loop and three-loop contributions to the ordinary phase free energy.}
\label{Orddiagram}
\end{figure}
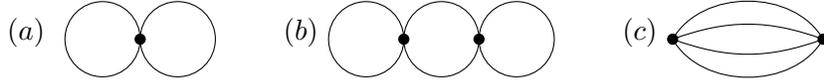

The order $\lambda_0$ term corresponds to the diagram $(a)$ in fig. \ref{Orddiagram}:
\begin{equation}
\begin{split}
\int_X &\left\langle\left(\phi^I\phi^I(X)\right)^2\right\rangle_0 = N(N+2){\rm Vol}(H^d) G_{\frac{d}{2}}(1)^2 \\
&= \frac{N(N+2)}{256\pi^4}{\rm Vol}(H^d)\left(1+\epsilon\, \xi+\epsilon^2 \left(\frac{\pi^2}{24}+\frac{1}{2}\xi^2\right)+\CO(\epsilon ^3)\right)
\end{split}
\end{equation}
where we have used \eqref{GdXX} for $G_{\frac{d}{2}}(1)$ and defined $\xi\equiv \gamma_E+\log(4\pi)$.  The order $\lambda_0^2$ term corresponds to the diagrams $(b)$ and $(c)$:
\begin{align}
\int_X\int_Y&\left\langle\left(\phi^I\phi^I(X)\right)^2 \left(\phi^J\phi^J(Y)\right)^2\right\rangle_0= 8N(N+2)^2 \CI_b+8N(N+2)\CI_c\\
&\CI_b\equiv G_{\frac{d}{2}}(1)^2 \int_{X, Y}G_{\frac{d}{2}}(X, Y)^2, \quad \CI_c\equiv  \int_{X, Y}G_{\frac{d}{2}}(X, Y)^4~.
\end{align}
To compute $\CI_b$, let's notice that the scalar propagator $G_\Delta(X, Y)$ can be thought as the matrix element of $(-\nabla^2+m^2)^{-1}$ in the $|X\rangle$ basis, where $m^2=\Delta(\Delta+1-d)$ \footnote{This picture is only valid for standard quantization.}. Then $\int_{X, Y}G_{\frac{d}{2}}(X, Y)^2$ is nothing but the trace of $(-\nabla^2+m^2)^{-2}$. The operator $(-\nabla^2+m^2)^{-2}$ is related to $(-\nabla^2+m^2)^{-1}$ by a mass derivative, or equivalently $\Delta$ derivative. Therefore, we have 
\begin{align}
\CI_b &= \frac{G_{\frac{d}{2}}(1)^2 {\rm Vol}(H^d)}{d-1-2\Delta}\left.\partial_\Delta G_\Delta(1)\right|_{\Delta\to\frac{d}{2}}\nonumber\\
&=\frac{{\rm Vol}(H^d)}{(4\pi) ^6}\left(\frac{2}{\epsilon}+1+3 \xi+\frac{\epsilon}{24} \left(6 \left(9 \xi ^2+6 \xi +2\right)-\pi ^2\right)+\CO(\epsilon^2 )\right)~.
\end{align}
For the evaluation of $\CI_c$, we can first put $Y$ at the origin of AdS by  using the $SO(d,1)$ symmetry and then  compute  the $X$ integral in the global coordinates with the substitution $\cosh r=\sigma$:
\begin{align}\label{ICc1}
\CI_c = {\rm Vol}(S^{d-1})  \left(\frac{ \Gamma(\frac{d}{2}-1)}{2^{\frac{d+2}{2}}\pi^{\frac{d}{2}}}\right)^4 \int_1^\infty d\sigma(\sigma^2-1)^{\frac{d}{2}-1} \left(\frac{1}{(\sigma-1)^{\frac{d}{2}-1}}- \frac{1}{(\sigma+1)^{\frac{d}{2}-1}}\right)^4~.
\end{align}
Expanding the last term of \eqref{ICc1} into monomials and using $\int_1^\infty\frac{d\sigma}{(\sigma-1)^a(\sigma+1)^b} =\frac{ \Gamma (1-a) \Gamma (a+b-1)}{2^{a+b-1}\Gamma (b)}$, we get 
\begin{align}\label{ICc2}
\CI_c = \frac{{\rm Vol}(H^d)}{1024 \pi ^6}\left( \frac{3}{ \epsilon }+\frac{7+54\xi}{12}+\frac{ \epsilon}{16} \left(2 \xi  (27 \xi +7)+7 \pi ^2-9\right)+\CO(\epsilon^2)\right)~.
\end{align}
Combining all the ingredients above leads to 
\begin{equation}\label{xx}
F_{\rm ord} =  N F_D \ + \frac{N(N+2)\textrm{Vol}(H^{d})}{1024 \pi^4} \bigg[ (1 + \epsilon \xi) \lambda - \frac{\lambda^2}{48 \pi^2}   ( 3 N+13+3(N+8) \xi )\bigg]~,
\end{equation}
which agrees with the calculation of  \cite{Giombi:2020rmc}. In \eqref{xx},  $\lambda$ is the renormalized coupling which takes the same form as in flat space, i.e. $\lambda_0 = \mu^\epsilon(\lambda + \frac{(N+8)\lambda^2}{8\pi^2\epsilon}+\cdots)$.
Plugging in the WF fixed point $\lambda_\star = \frac{8 \pi^2}{N+8} \epsilon   + \frac{24 (3N + 14) \pi^2}{(N + 8)^3} \epsilon^2+\CO(\epsilon^3)$ and expanding the AdS volume in $\epsilon$, the interacting part of the free energy becomes
\begin{align}\label{F2ordN}
F_{\rm ord} - N F_D = \frac{ N (N+2)}{96 (N+8)} \epsilon+\frac{5 N (N+2) (N^2+29N+132)}{576 (N+8)^3} \epsilon^2+\CO(\epsilon^3)~.
\end{align}

\subsection{The four-loop free energy}
\begin{figure}[t]
\centering
$(d)$\,\, \begin{tikzpicture} [baseline={(0,0)}]
\draw[color=black] (0,0) circle [radius=0.3]; 
\draw[color=black] (0.6,0) circle [radius=0.3]; 
\draw[color=black] (1.2,0) circle [radius=0.3]; 
\draw[color=black] (1.8,0) circle [radius=0.3]; 
\draw (0.3,0) node[vertex]{} ;\draw (0.9,0) node[vertex]{} ;\draw (1.5,0) node[vertex]{} ;
\end{tikzpicture} 
\qquad
$(e)$\,\,\begin{tikzpicture} [baseline={(0,0)}]
\draw[color=black] (0,0) circle [radius=0.5]; 
\draw[color=black] (0,0.8) circle [radius=0.3]; 
\draw[color=black] (0.69282,-0.4) circle [radius=0.3]; 
\draw[color=black] (-0.69282,-0.4) circle [radius=0.3]; 
\draw (0,0.5) node[vertex]{} ;
\draw (0.433013,-0.25) node[vertex]{} ;
\draw (-0.433013,-0.25) node[vertex]{} ;
\end{tikzpicture} 
\qquad
$(f)$\,\,\begin{tikzpicture} [baseline={(0,0)}]
\draw (0,0) node[vertex]{} to [out=-20, in=-160]  (2,0) node[vertex]{} to [out=120, in=60] (0,0) to [out=-60, in=-120] (2,0) to [out=160, in=20] (0,0);
\draw[color=black] (1,0.82) circle [radius=0.3]; \draw (1,0.52) node[vertex]{} ;
\end{tikzpicture} 
\qquad
$(g)$\,\,\begin{tikzpicture} [baseline={(0,0)}]
\draw[color=black] (0,0) circle [radius=0.5]; 
\draw (0,0.5) node[vertex]{} to (0.433013,-0.25) node[vertex]{} to (-0.433013,-0.25) node[vertex]{} to (0,0.5)  ;
\end{tikzpicture} 
\caption{The four-loop contributions to the ordinary phase free energy.}
\label{Orddiagram1}
\end{figure}
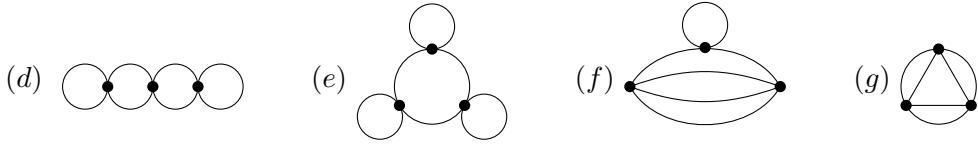

At the four-loop order, the free energy can be expressed as 
\begin{align}
F_{\rm 4-loop}^{\rm ord} = \lambda_0^3\left(\frac{N(N+2)^3}{4}\CI_d+ \frac{N(N+2)^3}{6}\CI_e+N(N+2)^2 \CI_f+\frac{N(N+2)(N+8)}{6}\CI_g\right)
\end{align}
where $\CI_d-\CI_g$ correspond to the diagrams $(d)-(g)$ in fig. \ref{Orddiagram1}. The symmetry factors are adapted from 
\cite{Gynther:2007bw}.

The evaluation of $\CI_d$ and $\CI_e$ is parallel to the evaluation of $\CI_b$. Here we just give the final result
\begin{align}
 \CI_d&= {\rm Vol}(H^d)\left.\left(\frac{\frac{1}{2}\partial_\Delta G_\Delta(1)^2}{d-1-2\Delta}\right)^2\right|_{\Delta\to\frac{d}{2}}=\frac{{\rm Vol}(H^d)}{4^7\pi ^8} \left(\frac{1}{ \epsilon
   ^2}+\frac{2 \xi +1}{ \epsilon }+\frac{24 \xi  (\xi +1)-\pi ^2+9}{12}+\CO(\epsilon)\right)~,\nonumber\\
 \CI_e&=\frac{1}{2} {\rm Vol}(H^d) G_{\frac{d}{2}}(1)^3\left. \left(\frac{\partial_\Delta }{d-1-2\Delta}\right)^2 G_\Delta(1)\right|_{\Delta\to\frac{d}{2}} = - \frac{{\rm Vol}(H^d) }{3\times 4^8 \pi^6}+\CO(\epsilon)~.
\end{align}

\subsubsection{$\CI_f$}
For the calculation of $\CI_f$, we find it useful to introduce the stereographic coordinates of AdS
\begin{align}\label{Stereo}
X^0 =  \frac{\cosh(t)}{\sinh(t)} , \quad X^a = \frac{\omega^a}{\sinh(t)}, 
\end{align}
where $t\ge 0$ and $\omega^a\in S^{d-1}$.
The metric in this coordinate system becomes $ds^2 = \Omega(x)^2 dx^2$ with $\Omega(x) \equiv  \frac{1}{\sinh(t)}$ and $dx^2\equiv dt^2+d\omega^2$. The propagator of a conformally coupled scalar with the Dirichlet boundary condition becomes
\begin{align}
G_{\frac{d}{2}} (X_1, X_2)& =  \frac{\CN_d}{(\Omega_1\Omega_2)^{\alpha}}\left(\frac{1}{(\cosh(t_1\!-\!t_2)\!-\!\omega_1\cdot\omega_2)^{\alpha}}- \frac{1}{(\cosh(t_1\!+\!t_2)\!-\!\omega_1\cdot\omega_2)^{\alpha}}\right), 
\end{align}
where $\alpha\equiv \frac{d}{2}-1$, $\CN_d= \frac{\Gamma(\alpha)}{4\pi(2\pi)^\alpha}$ and  $\Omega_i$ is a shorthand notation for $\Omega(x_i)= \frac{1}{\sinh(t_i)}$. 
In the stereographic coordinates, the diagram $(f)$ reads 
\begin{align}
\CI_f &=\int_{X_1, X_2, X_3} G_\frac{d}{2}(X_1, X_1)G_\frac{d}{2}(X_1, X_2) G_\frac{d}{2}(X_1, X_3) G_\frac{d}{2}(X_2, X_3)^3\nonumber\\
&= 2^4 \CN_d^5{\rm Vol}(H^d)G_{\frac{d}{2}}(1) \int_0^\infty dt_1 dt_2\frac{\sinh(\alpha t_1)\sinh^3(\alpha t_2)}{\sinh^2(t_1)\sinh^\epsilon(t_2)}\nonumber\\
&\times \int_{S^{d-1}}d^{d-1}\omega_1 d^{d-1}\omega_2\left(\frac{1}{(\cosh(t_1\!-\!t_2)\!-\!\omega_1\cdot\omega_2)^{\alpha}}-(t_2\leftrightarrow -t_2)\right)~,
\end{align}
where we have put $X_3$ at the origin of AdS.
We evaluate the sphere integrals in appendix \ref{harm}, cf. eq. \eqref{FDexact}, which gives 
\begin{align}
\CI_f &= - \frac{\Gamma(\alpha)^4 {\rm Vol}(H^d)}{2^{3d}\pi^{2d}\alpha^2}\int_0^\infty dt_1 dt_2\frac{\sinh(\alpha t_1)\sinh^3(\alpha t_2)}{\sinh^2(t_1)\sinh^\epsilon(t_2)}\left(e^{-\alpha|t_1-t_2|} - e^{-\alpha(t_1+t_2)}\right)\nonumber\\
&= - \frac{\Gamma(\alpha)^4 {\rm Vol}(H^d)}{2^{3d-1}\pi^{2d}\alpha^2}\int_0^\infty dt_2 f_\epsilon (t_2)~,\\
f_{\epsilon}(t_2) &\equiv \frac{\sinh^4(\alpha t_2)}{\sinh^\epsilon(t_2)}\int_{t_2}^\infty dt_1 \frac{\sinh(\alpha t_1)e^{-\alpha t_1}}{\sinh^2(t_1)}+\frac{\sinh^3(\alpha t_2) e^{-\alpha t_2}}{\sinh^\epsilon(t_2)}\int_0^{t_2} dt_1 \frac{\sinh(\alpha t_1)^2}{\sinh^2(t_1)}~.
\end{align}
The $t_1$ integrals in $f_{\epsilon}(t_2)$ admit the following series expansions, which are obtained by  expanding the integrands in $e^{-t_1}$
\begin{align}
&\int_{t_2}^\infty dt_1 \frac{\sinh(\alpha t_1)e^{-\alpha t_1}}{\sinh^2(t_1)} = \sum_{n=0}^\infty\left(e^{-2(n+1)t_2}-\frac{n+1}{n+\alpha+1}e^{-2(n+\alpha+1)t_2}\right)~,\nonumber\\
&\int_0^{t_2} dt_1 \frac{\sinh(\alpha t_1)^2}{\sinh^2(t_1)}=\frac{1-\alpha\pi\cot(\alpha\pi)}{2}+\sum_{n=0}^\infty\left(e^{-2(n+1)t_2}-\sum_{\pm }\frac{n+1}{2(n+1\pm \alpha)}e^{-2(n+1\pm \alpha) t_2}\right)~.
\end{align}
These series expansions allow us to rewrite $f_{\epsilon}(t_2)$ as 
\begin{align}
f_{\epsilon}(t_2)& = \frac{\sinh^3(\alpha t_2)}{\sinh^\epsilon(t_2)} \left[\frac{1-\alpha\pi\cot(\alpha\pi)}{2} e^{-\alpha t_2} + \frac{\alpha}{2}\sum_{n\ge 0} \left( \frac{ e^{-(2n+2+\alpha) t_2}}{n+1+ \alpha}- \frac{ e^{-(2n+2- \alpha) t_2}}{n+1- \alpha}\right)\right]
\end{align}
We can evaluate the $t_2$ integral for each term in the bracket using 
\begin{align}\label{s3}
\Phi_\epsilon(\kappa)&\equiv \int_0^\infty dt\,\frac{\sinh^3(\alpha t)e^{-\kappa t}}{\sinh^\epsilon(t)}\nonumber\\
&=\frac{ \Gamma (1-\epsilon )}{2^{4-\epsilon }} \left(-\frac{\Gamma \left(\frac{6-\epsilon +2 \kappa }{4} \right)}{\Gamma
   \left(\frac{10-5 \epsilon +2 \kappa }{4} \right)}+\frac{3 \Gamma \left(\frac{2+\epsilon +2 \kappa
   }{4} \right)}{\Gamma \left(\frac{6-3 \epsilon +2 \kappa }{4} \right)}-\frac{3 \Gamma \left(\frac{3
   \epsilon +2 \kappa -2}{4}\right)}{\Gamma \left(\frac{2-\epsilon +2 \kappa }{4} \right)}+\frac{\Gamma
   \left(\frac{5 \epsilon +2 \kappa -6}{4} \right)}{\Gamma \left(\frac{\epsilon +2 \kappa
   -2}{4} \right)}\right)~,
\end{align}
and then we obtain a series representation for the integral $\int_0^\infty dt_2 f_\epsilon (t_2)$
\small
\begin{align}\label{f2n}
\int_0^\infty dt_2 f_\epsilon (t_2) &= \frac{1\!-\!\alpha\pi\cot(\alpha\pi)}{2} \Phi_\epsilon(\alpha)+\sum_{n\ge 0}\hat\gamma_n(\epsilon),\quad
 \hat\gamma_n(\epsilon)\equiv \frac{\alpha}{2}\left( \frac{\Phi_\epsilon(2n\!+\!2\!+\!\alpha)}{n+1+ \alpha}- \frac{ \Phi_\epsilon(2n\!+\!2\!-\!\alpha)}{n+1- \alpha}\right)~.
 \end{align}
 \normalsize
 The coefficients $ \hat\gamma_n(\epsilon)$ satisfy the following two properties: (\rom{1}) both $ \hat\gamma_0(\epsilon)$ and $ \hat\gamma_1(\epsilon)$ have a pole at $\epsilon=0$ while all  $ \hat\gamma_{n\ge 2}(\epsilon)$ are regular at $\epsilon=0$, and (\rom{2}) the leading large $n$ asymptotic behavior of $ \hat\gamma_n(\epsilon)$ is $n^{\epsilon-6}$. Therefore, up to corrections of order $\epsilon$, we can replace the sum of $ \hat\gamma_n(\epsilon)$ for $n\ge 2$ in \eqref{f2n} by $\sum_{n\ge 2}\hat\gamma_n(0) = \frac{\pi^2-10}{32}$~. 
 
Combining all the ingredients above, we find the $\epsilon$-expansion of the diagram $(f)$ to be 
\begin{align}
\CI_f & = - \frac{\Gamma(\alpha)^4 {\rm Vol}(H^d)}{2^{3d-1}\pi^{2d}\alpha^2}\left(\frac{1-\alpha\pi\cot(\alpha\pi)}{2} \Phi_\epsilon(\alpha)+ \hat\gamma_0(\epsilon)+ \hat\gamma_1(\epsilon)+ \frac{\pi^2-10}{32}+\CO(\epsilon) \right)\nonumber\\
&= {\rm Vol}(H^d)\left(\frac{3}{2^{15}\pi ^8 \epsilon ^2}+\frac{17+24 \xi }{2^{17} \pi ^8 \epsilon }+\frac{\frac{229}{12}-3 \pi
   ^2+2 \xi  (12 \xi +17)}{2^{17} \pi ^8}+\CO\left(\epsilon \right)\right)~.
\end{align}

\subsubsection{$\CI_g$}
We also compute $\CI_g$ in the coordinate system \eqref{Stereo}. By using the isometry, we move one vertex to the origin of AdS, and then $\CI_g$ becomes
\begin{align}\label{Igsome}
\CI_g&= 4\CN_d^6 {\rm Vol}(H^d)\int_{0}^\infty dt_1 dt_2\int d^{d-1}\omega_1 d^{d-1}\omega_2 \frac{(\cosh(2\alpha t_1)-1)(\cosh(2\alpha t_2)-1)}{\sinh^\epsilon(t_1)\sinh^\epsilon(t_2)}\nonumber\\
&\times \left(\frac{1}{(\cosh(t_1\!-\!t_2)\!-\!\omega_1\cdot\omega_2)^{\alpha}}- \frac{1}{(\cosh(t_1\!+\!t_2)\!-\!\omega_1\cdot\omega_2)^{\alpha}}\right)^2~.
\end{align}
We first analyze  how $(\cosh (t_{1}\pm t_2)-\omega_1\cdot\omega_2)^{-2\alpha}$ contributes to $\CI_g$. As shown in appendix \ref{harm} , the angular integral of this term admits the following Fourier representation 
\begin{align}\label{nuint}
\int \frac{d^{d-1}\omega_1 d^{d-1}\omega_2}{(\cosh (t_{1}\pm t_2)-\omega_1\cdot\omega_2)^{2\alpha}} =\int_{\mathbb R}d\nu\, \CK(\nu) e^{i\nu (t_{1}\pm t_2)}
\end{align}
where the function $\CK(\nu)$ is defined in \eqref{FD}.
 It  allows us to perform the $t_1$ and $t_2$ integrals analytically
 \small
\begin{align}
\psi(\nu)\equiv \int_0^\infty dt \frac{\cosh(2\alpha t)-1}{\sinh^\epsilon(t)} e^{i\nu t} =\frac{ \Gamma (d\!-\!3 )}{2^{d-2}} \left(\frac{\Gamma \left(1\!-\!\frac{i \nu }{2}\right)}{\Gamma \left(2\!-\!\epsilon \!-\!\frac{i \nu }{2}\right)}+\frac{\Gamma \left(\epsilon\!-\!1 \!-\!\frac{i \nu
   }{2}\right)}{\Gamma \left(-\frac{i \nu }{2}\right)}-\frac{2 \Gamma
   \left(\frac{\epsilon -i \nu }{2}\right)}{\Gamma \left(1\!-\!\frac{\epsilon }{2}\!-\!\frac{i \nu }{2}\right)}\right)~,
\end{align}
\normalsize
which is convergent for $1<\epsilon<3$.
Then the full contribution of $(\cosh (t_{1}\pm t_2)-\omega_1\cdot\omega_2)^{-2\alpha}$ can be concisely expressed as 
\begin{align}\label{Ig1def}
1<\epsilon<3: \quad \CI^{(1)}_g =  4\CN_d^6 {\rm Vol}(H^d)\int_{\mathbb R}d\nu\,( \psi(\nu)+\psi(-\nu))\psi(\nu)\CK(\nu) ~,
\end{align}
Next, we  analytically continue $ \CI^{(1)}_g$ to $\epsilon\in (0, 1)$. Although the $\nu$ integral remains convergent for $0<\epsilon<1$, the analytical continuation is quite non-trivial due to the poles of $\psi(\pm\nu)$ at $\nu=\pm 2i(1-\epsilon)$. As $\epsilon$ decreases from $1^+$ to $1^-$,  these poles cross the $\nu$ contour. Therefore, we must manually subtract the residues of the poles entering the upper half-plane and add the residues of those exiting it.
Taking into account of these residues, we find the proper analytical continuation of $\CI_g^{(1)}$ near 4 dimensions to be
\begin{align}
0<\epsilon<1: \quad \CI^{(1)}_g &= {\rm Vol}(H^d)\bigg( 2\CN_d^6 \int_{\mathbb R}d\nu\,( \psi(\nu)\!+\!\psi(-\nu))^2\CK(\nu)\nonumber\\
&+ \frac{3}{2^{14} \pi ^8 \epsilon ^2}+\frac{5/3\!+\!12 \xi}{2^{15} \pi ^8 \epsilon }+\frac{4 \xi  (18 \xi +5)+11 \pi ^2-38}{3\times 2^{16} \pi ^8}+\CO(\epsilon)\bigg)
\end{align}
where we have performed the small $\epsilon$ expansion for the residue part and used $\CK(\nu) = \CK(-\nu)$. The terms corresponding to pole crossing are crucial to get a finite free energy after renormalization.

 For the remaining $\nu$ integral, we notice that the function $\psi(\nu)+\psi(-\nu)$ has poles at $ \pm i\epsilon$ and $\pm 2i\epsilon$, which pinch the contour $\mathbb R$ in the $\epsilon\to 0$ limit. It forbids us to take the $\epsilon\to 0$ limit inside the integral. 
As an illustration of our strategy of  dealing with such singularities, let's consider an  exactly solvable integral  with the same feature, e.g. $\int_{\mathbb R}\frac{dx }{x^2+\epsilon^2}=\frac{\pi}{\epsilon}$. Taking $\epsilon\to0$ of the integrand leads to the $\frac{1}{x^2}$ singularity at $x=0$. Instead, we can move the contour to $\mathbb R+i$ and then safely expand the integrand in $\epsilon$,
\begin{align}\label{dd}
\int_{\mathbb R+i}\frac{dz}{z^2+\epsilon^2} = \int_{\mathbb R}\frac{dx}{(x+i)^2+\epsilon^2} = \sum_n(-)^n \epsilon^{2n}  \int_{\mathbb R}dx\frac{1}{(x+i)^{2(n+1)}} = 0~.
\end{align}
Since the singularity at $i\epsilon$ crosses the contour during the contour deformation, we need to add $2\pi i \, {\rm Res}_{z=i\epsilon}\frac{1}{z^2+i\epsilon} = \frac{\pi}{\epsilon}$, which together with \eqref{dd} reproduces the correct result.
Coming back to the $\nu$ integral in $\CI^{(1)}_g$, we move the contour to $\gamma=\mathbb R+\frac{i}{4}$ which allows us to take the $\epsilon\to 0$ limit inside the integral. The new integral along $\gamma$ appears to be of order $\epsilon$ because $\CK$ has a $\frac{1}{\epsilon}$ pole and the sum of $\psi(\pm \nu)$  is of order $\epsilon$. Due to the contour deformation, we pick up the residues at $\nu=i\epsilon$ and $\nu= 2i\epsilon$. Altogether, we find the $\epsilon$-expansion of  $\CI_g^{(1)}$ near 4 dimensions to be 
\begin{align}\label{CIg1}
\CI_g^{(1)}={\rm Vol}(H^d)\bigg(\frac{5}{2^{14} \pi ^8 \epsilon ^2}+\frac{5 (4 \xi +1)}{2^{15}\pi ^8 \epsilon }+\frac{60 \xi  (2 \xi +1)+21 \pi
   ^2-41}{3\times 2^{16}\pi ^8}+\CO(\epsilon )\bigg )~.
\end{align}

For the  cross term in \eqref{Igsome}, the angular integrals follow from \eqref{ddtK}.
\begin{align}
\int\frac{d^{d-1}\omega_1d^{d-1}\omega_2}{(\cosh (t_1-t_2)-\omega_1\cdot\omega_2)^{\alpha}(\cosh (t_1+t_2)-\omega_1\cdot\omega_2)^{\alpha}} = \frac{(2\pi)^d}{\Gamma(\frac{d}{2})^2} \,
 \frac{ F\left(\alpha,2\alpha;\alpha+1, e^{-2t}\right)}{e^{2\alpha t }}~,
\end{align}
where $t=t_1$ when $t_1\ge t_2$ and $t=t_2$ when $t_1<t_2$. The remaining $t$ integrals become
\begin{align}
\CI^{(2)}_g &=  -\frac{2^4(2\pi)^d}{\Gamma(\frac{d}{2})^2} \CN_d^6 {\rm Vol}(H^d)\int_0^\infty dt_1  g_\epsilon(t_1)~,\nonumber\\
g_\epsilon(t_1)&=\frac{\cosh(2\alpha t_1)-1}{\sinh^\epsilon(t_1)}  \frac{ F\left(\alpha,2\alpha;\alpha+1, e^{-2t_1}\right)}{e^{2\alpha t_1 }}\int_0^{t_1} dt_2  \frac{\cosh(2\alpha t_2)-1}{\sinh^\epsilon(t_2)}~,
\end{align}
where in the $\epsilon\to 0$ limit $g_\epsilon$ reduces to $g_0(t) = \frac{1}{4}(1-e^{-2t})(\sinh(2t)-2t)$. 
We will derive  the divergent and the finite parts of $\int dt_1 g_\epsilon(t_1)$ by using the subtraction method.  More explicitly, we choose some function $\hat g_\epsilon(t_1)$, such that the integral of 
$g_\epsilon  - \hat g_\epsilon$ is finite for $0\le \epsilon<\epsilon_c$ (the explicit value of $\epsilon_c$ is not important), and the integral of $\hat g_\epsilon$ can be evaluated analytically for any $\epsilon$ (and hence we can do analytical continuation in $\epsilon$ easily). Once such a $\hat g_\epsilon$ is found, the integral of $g_\epsilon$ can be determined up to errors of order $\epsilon$ 
\begin{align}\label{vvvv}
 \int_0^\infty dt_1  g_\epsilon(t_1) =  \int_0^\infty d t_1 \left( g_0(t_1)-\hat g_0(t_1)\right)+ \int_0^\infty dt_1  \hat g_\epsilon(t_1)+\CO(\epsilon)~.
\end{align}
Then we explain how to obtain a simple  $\hat g_\epsilon$ by modifying $g_\epsilon$. The first step is to replace the hypergeometric function in $g_\epsilon$ by $1+\frac{2\alpha^2}{1+\alpha}e^{-2t_1}$. Second,  expand $\frac{\cosh(2\alpha t_2)-1}{\sinh^\epsilon(t_2)}$ as a series of $e^{-t_2}$
\begin{align}
\frac{\cosh(2\alpha t_2)-1}{\sinh^\epsilon(t_2)} = 2^{\epsilon-1}\sum_{n\ge 0} \left( e^{-(2n+2\alpha-\epsilon)t_2}+e^{-(2n-2\alpha-\epsilon)t_2}-2e^{-(2n-\epsilon)t_2}\right)~.
\end{align}
For each $n$, the $t_2$ integral can be carried out easily. After the $t_2$ integral, we keep terms proportional to $e^{-(2n+2\alpha-\epsilon)t_1}, e^{-(2n-2\alpha-\epsilon)t_1}, e^{-(2n-\epsilon)t_1}$ for $n=0, 1$, as well as the $t_1$ independent piece. These two steps fix a choice of $\hat g_\epsilon$. It is easy to check that $\hat g_\epsilon$ exactly removes the problematic large $t_1$ behavior of $g_\epsilon$. Also since $\hat g_\epsilon(t_1)$ is just a linear combination of $e^{\# t_1}/\sinh^\epsilon(t_1)$, its integral can be computed analytically.  Using \eqref{vvvv}, we find
\begin{align}\label{CIg2}
\CI_g^{(2)} = {\rm Vol}(H^d)\bigg(\frac{1}{2^{12} \pi ^8 \epsilon^2}+\frac{2 \xi +1}{2^{12} \pi ^8\epsilon}+\frac{8 \xi  (\xi +1)+ \pi ^2+5/3}{2^{14} \pi ^8} +\CO(\epsilon) \bigg)~.
\end{align}
Combining \eqref{CIg1} and \eqref{CIg2}, we finally obtain the $\epsilon$-expansion of the diagram $(g)$
\begin{align}\label{CIg}
\CI_g = {\rm Vol}(H^d)\bigg( \frac{9}{2^{14}\pi ^8 \epsilon ^2}+\frac{36 \xi +13}{2^{15}\pi ^8 \epsilon }+\frac{72 \xi ^2+52 \xi +11 \pi ^2-7}{2^{16}
   \pi ^8}+\CO(\epsilon)\bigg )~.
\end{align}

\subsubsection{Curvature counterterms}
The renormalization of interacting scalar theories on a curved manifold typically requires curvature counterterms. For the O$(N)$ model, we need to add \cite{Brown:1980qq,Hathrell:1981zb,Jack:1983sk,Jack:1990eb}
\begin{align}\label{cct}
S_{\rm c.t} = \int d^d x\sqrt{g}\left(\frac{1}{2}\eta_0 H \phi_0^I \phi_0^I + a_0 W^2 +b_0 E+c_0 H^2\right)
\end{align}
to the action \eqref{Sor}. Here we have explicitly included the subscript ``0'' to emphasize that all the couplings and fields are bare quantities. In \eqref{cct}, $W^2$ denotes the square of the Weyl tensor, $E$ is the Euler density and $H=\frac{{\cal R}}{d-1}$ is proportional to the Ricci scalar ${\cal R}$. For AdS$_d$ of radius 1, the Weyl tensor vanishes, and Euler density is equal to $d(d-1)(d-2)(d-3)$. 

The renormalization of these curvature couplings on a general curved manifold have been calculated in \cite{Brown:1980qq,Hathrell:1981zb,Jack:1983sk}. In our case and at the order of interest, only the Euler density term is relevant and the corresponding fixed point is  $b_\star = - 5\frac{b_{41}}{\epsilon}\lambda_\star^4+\cdots$, where 
\begin{align}
 b_{41} = -\frac{N(N+2)(3N+14)}{240(4\pi)^{10}}~.
\end{align}
The free energy counterterm is thus 
\begin{align}
F_{\rm c.t} = b_\star \int d^d x\sqrt{g} E =- \frac{N(N+2)(3N+14)}{384(N+8)^4}\epsilon^3~.
\end{align}

\subsubsection{Free energy at the WF fixed point}

Adding diagrams $(a) - (g)$ together with the curvature counterterm, we obtain the full 4-loop free energy of the O$(N)$ model in AdS at the WF fixed point
\small
\begin{align}\label{F4}
F_{\rm ord} &=N F_D +\frac{ N (N+2)}{96 (N+8)} \epsilon+\frac{5 N (N+2) (N^2+29N+132)}{576 (N+8)^3} \epsilon^2\nonumber\\
&+\frac{N (N+2)\epsilon^3}{6912 (N+8)^5}\bigg(8 \left(7+\pi ^2\right) N^4+\left(2409+184 \pi ^2\right) N^3+6 N^2 \left(-720 \zeta (3)+7309+272 \pi ^2\right) \nonumber\\
&+4 N\left(-13392 \zeta (3)+72265+1792 \pi ^2\right)+32 \left(-4752 \zeta (3)+19983+448 \pi ^2\right)\bigg) +\CO(\epsilon^4)~.
\end{align}
\normalsize
As a consistency check, we expand  $F_{\rm ord}$ for large $N$ 
\begin{align}\label{F4L}
F_{\rm ord} = N \left(F_D + \frac{\epsilon}{96}+ \frac{5\epsilon^2}{576}+ \frac{(\pi^2+7)\epsilon^3}{864}+\CO(\epsilon^4)\right)+\CO(N^0)~.
\end{align}
Eq.\eqref{F4L} agrees with the large $N$ analysis in \cite{Giombi:2020rmc}, where it is shown that in the leading order of the large $N$ expansion, $F_{\rm ord}$ should be equal to the one-loop partition function of $N$ scalar fields of boundary dimension $d-2$:
\begin{align}
N F_1(d-2)  = N F_D+\frac{N{\rm Vol}(H^{d})}{2(4\pi)^{\frac{d}{2}}}\int^{d/2-2}_{0}\,d\hat\Delta\frac{(2\hat\Delta+1)\Gamma(\hat\Delta+\frac{d}{2})\Gamma(1-\frac{d}{2})}{\Gamma(2-\frac{d}{2}+\hat\Delta)}~,
\end{align}
where we have used  \eqref{F1mD}. To find the $\epsilon$ expansion of the $\hat\Delta$ integral, 
we first make the change of variable $\hat\Delta\to -\frac{\epsilon}{2} x$ and then expand the integrand in $\epsilon$ to the order of $\epsilon^3$. Performing the $x$ integral for each term in the expansion recovers exactly \eqref{F4L}.

To derive $\tilde s$ as defined in \eqref{tilde-s}, we also need the sphere free energy $F_{S^{d}}$ of the O($N$) model at the WF fixed point. This quantity was computed in \cite{Fei:2015oha} to the order of $\epsilon^4$ and the result can be summarized as 
\begin{align}\label{Fsphere}
F_{S^d} = N F_{\rm conf} -\frac{N (N+2) \epsilon ^2}{288 (N+8)^2}-\frac{N (N+2) \left(13 N^2+370 N+1588\right) \epsilon ^3}{3456
   (N+8)^4}+\CO\left(\epsilon^4\right)
\end{align}
where $F_{\rm conf} $ is the sphere free energy of a conformally coupled scalar given by \eqref{Fs}.

Combining \eqref{F4} and \eqref{Fsphere}, we obtain $\ts$ for the ordinary boundary condition of the O$(N)$ model in $d=4-\epsilon$ dimensions to the order of $\epsilon^3$
\begin{align}\label{Ford4}
\widetilde s^{\,\rm ord}_{O(N)}(d) &=-N(0.00761211 +0.0097764 \epsilon+0.00720205 \epsilon ^2 +0.0041371 \epsilon ^3)\nonumber\\
&+\frac{N (N+2) \epsilon }{96 (N+8)}+\frac{N (N+2) \left(5 N^2+146 N+668\right) \epsilon ^2}{576 (N+8)^3}\\
&-\frac{ N(N+2)\epsilon ^3}{6912
   (N+8)^5} \bigg(\left(\pi ^2-56\right) N^4+2 \left(52 \pi ^2-1211\right) N^3+24 N^2 \left(180 \zeta (3)-1847+76 \pi ^2\right)\nonumber\\
   &+8 N
   \left(6696 \zeta (3)-36701+1408 \pi ^2\right)+128 \left(1188 \zeta (3)-5095+176 \pi ^2\right)\bigg)+\CO(\epsilon^4)~,\nonumber
\end{align}
where the first line comes from $N(F_D- F_{\rm conf}) = \frac{N}{2}(F_D-F_N)$, which is given by \eqref{FDmN}. In particular, for the Ising model, i.e. $N=1$, \eqref{Ford4} becomes
\begin{align}
\widetilde s_{\rm Ising}^{\,\rm ord} (d)=-0.0076121-0.00630417 \epsilon -0.00135072 \epsilon^2 -0.00128777\epsilon^3+\CO(\epsilon^4)~.
\end{align}

\subsubsection{Pad\'e approximation of the free energy}

In three dimensions, $\widetilde s$ is proportional to the boundary central charge, i.e. $\widetilde s = \frac{\pi}{6}c$, while 
in two dimensions,  $\widetilde s$ goes to the logarithm of the boundary $g$ function.  For the ordinary boundary condition of  the 2$d$ Ising model, the $g$ function is equal to 1 \cite{Dorey:2009vg} \footnote{The ordinary boundary condition corresponds to the free boundary condition in \cite{Dorey:2009vg}.} and hence we have $\widetilde s_{\rm Ising}^{\,\rm ord} (2) = 0$. We can try the two-sided $[n,m]$ Pad\'e for $\widetilde s_{\rm Ising}^{\,\rm ord} (4-\epsilon)$ with total order $n+m =4$.  
However,  the $[0,4]$ Pad\'e is not consistent with with the boundary value at $d=2$, and all the rest Pad\'e approximants appear to have a pole for $2<d<4$. For this reason, we consider instead  $\exp(\widetilde s_{\rm Ising}^{\,\rm ord} )$ and its [0,4] two-sided Pad\'e turns out to be regular for all real $d$ 
\begin{align}
\text{Pad\'e}_{[0,4]}: \quad \exp(\widetilde s_{\rm Ising}^{\,\rm ord} )= \frac{0.9924}{1+0.006304 \epsilon+0.001371 \epsilon^2+0.001296 \epsilon^3-0.002253 \epsilon^4}~.
\end{align}
In three dimension, it leads to the boundary central charge  $c_{\rm Ising}^{\rm ord} \approx -0.0273$. This appears to be not so close to the  fuzzy sphere result $c_{\rm fuzzy}\approx -0.0159(5)$ obtained recently in \cite{Zhou:2024dbt}, although it is much closer than the large $N$ estimate  $c|_{\rm large N} =-\frac{1}{16}\approx -0.0625$ \cite{Krishnan:2023cff}.   On the other hand, if we drop the $\epsilon^3$ corrections in the free energy, then the two-sided [1,2]  Pad\'e approximant of $\widetilde s_{\rm Ising}^{\,\rm ord}$ is well-defined for $2\le d<4$ and yields  $c_{\rm Ising}^{\rm ord} \approx -0.012$, which is closer to the  fuzzy sphere result. 

Let us mention that the surface defect calculation  in \cite{Giombi:2023dqs, Diatlyk:2024ngd} appears to be consistent with our 4-loop estimate $c_{\rm Ising}^{\rm ord} \approx -0.0273$. 
More explicitly,  \cite{Giombi:2023dqs, Diatlyk:2024ngd} considered the O$(N)$ model in $d=4-\epsilon$ dimensions with a surface defect $\CD = \int d^2 x\phi^I\phi^I$, and computed the defect $b$-anomaly by wrapping $\CD$ on a sphere \footnote{The leading term of the  $\epsilon$-expansion of $b$ was derived in \cite{Giombi:2023dqs} and the subleading term was calculated in \cite{Diatlyk:2024ngd}.}
\begin{align}\label{bON}
b_{O(N)}= -\frac{27N\epsilon^3}{(N+8)^3}- \frac{27N(N+2)(52-N)\epsilon^4}{4(N+8)^5}+\CO(\epsilon^5)~.
\end{align}
In $d=3$ dimension, the defect becomes an interface and factorizes into a pair of ordinary boundary universality classes \cite{Bray:1977fvl, Krishnan:2023cff}. So $b_{O(N)}$ is twice of the boundary central charge of the ordinary boundary universality class in 3$d$. For $N=1$ and $\epsilon=2$, the defect becomes a mass deformation in the full spacetime and triggers an RG flow from the Ising mimimal model to a massive phase. So $b_{\rm Ising} = -\frac{1}{2}$ in 2$d$. With the exact value in two dimensions, we can apply two-sided Pad\'e to \eqref{bON} for $N=1$. For example, the [1,1] Pad\'e approximant reads
\begin{align}
\text{Pad\'e}_{[1,1]}:\quad b_{\rm Ising} = -\frac{ (44+29 \epsilon)\epsilon^3}{6 (198+37 \epsilon)} ~,
\end{align}
leading to the boundary central charge $c_{\rm Ising}^{\rm ord} \approx-0.0259$ in 3$d$. It is in good agreement with our estimate above obtained from a four-loop calculation in AdS. 

For higher $N$, only the one-sided [1,2] Pad\'e approximant of $\widetilde s_{\rm Ising}^{\,\rm ord} $ is available, based on which we obtain the estimates  $c_{O(2)}^{\rm ord} \approx -0.055$ and $c_{ O(3)}^{\rm ord} \approx  -0.073$. 
Similarly, the one-side [0,1] Pad\'e approximant of $b_{O(N)}$ for the surface defect yields  $c_{O(2)}^{\rm ord} \approx-0.054$ and $c_{ O(3)}^{\rm ord} \approx -0.062$.


\subsection{The $\phi^2$ one-point function}

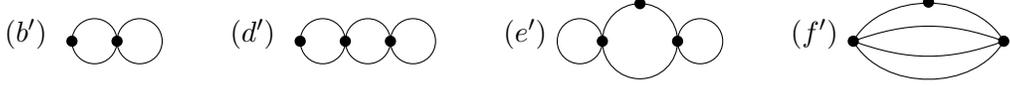
\begin{figure}[t]
\centering
$(b')$\,\, \begin{tikzpicture} [baseline={(0,0)}]
\draw[color=black] (0.6,0) circle [radius=0.3]; 
\draw[color=black] (1.2,0) circle [radius=0.3]; 
\draw (0.3,0) node[vertex]{} ;\draw (0.9,0) node[vertex]{} ;
\end{tikzpicture} 
\qquad
$(d')$\,\, \begin{tikzpicture} [baseline={(0,0)}]
\draw[color=black] (0.6,0) circle [radius=0.3]; 
\draw[color=black] (1.2,0) circle [radius=0.3]; 
\draw[color=black] (1.8,0) circle [radius=0.3]; 
\draw (0.3,0) node[vertex]{} ;\draw (0.9,0) node[vertex]{} ;\draw (1.5,0) node[vertex]{} ;
\end{tikzpicture} 
\qquad
$(e')$\,\,\begin{tikzpicture} [baseline={(0,0)}]
\draw[color=black] (0,0) circle [radius=0.5]; 
\draw[color=black] (0.8,0) circle [radius=0.3]; 
\draw[color=black] (-0.8,0) circle [radius=0.3]; 
\draw (0,0.5) node[vertex]{} ;
\draw (0.5,0) node[vertex]{} ;
\draw (-0.5,0) node[vertex]{} ;
\end{tikzpicture} 
\qquad
$(f')$\,\,\begin{tikzpicture} [baseline={(0,0)}]
\draw (0,0) node[vertex]{} to [out=-20, in=-160]  (2,0) node[vertex]{} to [out=120, in=60] (0,0) to [out=-60, in=-120] (2,0) to [out=160, in=20] (0,0); \draw (1,0.52) node[vertex]{} ;
\end{tikzpicture} 
\caption{Corrections to the $\phi^2$ one-point function.}
\label{phi2diagram}
\end{figure}

The AdS set-up also provides a convenient way to compute the one-point function of primary operators directly because it is simply a constant. More explicitly, let $\CO$ be a primary operator in AdS with scaling dimension $\Delta_\CO$ and one-point function $\langle \CO\rangle_{H^d} = c_\CO $. After mapping it back to the flat space with boundary, 
we get $\langle \CO(z)\rangle_{\rm bdry} = \frac{a_\CO}{(2z)^{\Delta_\CO}}$ where $z$ is the distance to the boundary and $a_\CO = 2^{\Delta_\CO}c_\CO$ is the one-point coefficient of $\CO$ in the BCFT convention.

In this subsection, we derive the one-point function of the O($N$) invariant  operator $\phi^2$. The leading order contribution is simply a tadpole diagram, i.e. $N G_{\frac{d}{2}}(1)$. Up to the $\lambda^2$ order, the relevant diagrams are shown in fig. \ref{phi2diagram}. 
A useful trick is to compute the integrated one-point function, which is equal to $\langle \phi_0^2\rangle$ multiplied by the AdS volume. On the other hand, the integrated one-point function can be related to the free energy. For example, the diagram $(f')$ in fig. \ref{phi2diagram} is the same as the diagram $(f)$ in fig. \ref{Orddiagram1} with one tadpole subdiagram stripped off. Based on  this observation,  the  one-point function of  the bare   $\phi^2_0$ operator can be expressed as 
\begin{align}
\left\langle \phi^2_0 \right\rangle =N G_{\frac{d}{2}}(1)+ \frac{-\lambda_0 N(N+2) \CI_b + \lambda_0^2N(N+2)^2(\CI_d+\CI_e)+2\lambda_0^2 N(N+2)\CI_f}{G_{\frac{d}{2}}(1) {\rm Vol}(H^d)}~.
\end{align}

In flat space, the $\phi_0^2$ operator is renormalized by a multiplicative factor, i.e. $ [\phi^2]_R = Z_{\phi^2}^{-1} \phi_0^2$,
where $Z_{\phi^2} =1-\frac{\lambda  (N+2)}{8 \pi ^2 \epsilon }+\frac{\lambda ^2 (N+2) (5 \epsilon -12)}{256 \pi ^4 \epsilon ^2}+\CO(\lambda^3)$ is the wavefunction renormalization. This is {\it not} generally true in  a curved manifold $\CM$ because there is another dimension 2 operator, the Ricci scalar ${\cal R}_\CM$. It can mix with the $\phi_0^2$ operator along the RG flow \cite{Hathrell:1981zb}. So the renormalized $\phi_0^2$ operator on $\CM$ should be
$[\phi^2]_{\CM} \equiv [\phi^2]_R  +\kappa \mu^{-\epsilon}\, {\cal R}_{\CM}$, where $\kappa$ is a function of the coupling and is independent of $\CM$.
To the order we are concerned with, the curvature term is not necessary to define a finite $\phi^2$ operator \cite{Hathrell:1981zb}. However, the curvature term is required to construct a primary $\phi^2$ operator at the WF fixed point. One can fix the finite counter term $\kappa$ by requiring that $\langle [\phi^2]_{H^d}[\phi^2]_{H^d}\rangle $ takes  the form of a conformal two-point function on the hyperbolic space at the fixed point. A more efficient strategy, based on the $\CM$-independence of $\kappa$, is to compute the one-point function on a round sphere and impose the condition   $\langle [\phi^2]_{S^d}\rangle =0 $ at the fixed point.
We carry out this calculation in appendix  \ref{sphereagain}, and then find the correct primary operator on AdS to be
 \begin{align}
[\phi^2]_{H^d} \equiv [\phi^2]_R  -\frac{N(N+2)\epsilon^2}{576\pi^2(N+8)^2} \mu^{-\epsilon}\, {\cal R}_{H^d}
\end{align}
which holds up to the $\epsilon^2$ order at the WF fixed point.
We further normalize this operator by using the bulk two-point function coefficient $\CN_{\phi^2}$ in flat space, defined as 
\begin{align}
\left\langle [\phi^2]_R (x)  [\phi^2]_R (0)\right\rangle_{\mathbb R^d} = \frac{\CN_{\phi^2}^2}{x^{2\Delta_{\phi^2}}}~.
\end{align}
The explicit form of  $\CN_{\phi^2}$ up to order $\epsilon^2$ is given by eq. (B.13) of \cite{Diatlyk:2024ngd}. Altogether, at the WF fixed point, we find the normalized one-point function to be 
\begin{align}\label{aphi2}
\left\langle \CN_{\phi^2}^{-1} [\phi^2]_{H^d}\right\rangle =2^{-\Delta_{\phi^2}}a^{\rm ord}_{\phi^2}, \quad a^{\rm ord}_{\phi^2} = -\sqrt{\frac{N}{2}}\left(1+\frac{(N+2)^2 \epsilon ^2}{8(N+8)^2}\right)+\CO\left(\epsilon ^3\right)~.
\end{align}

The one-point function coefficient of $\phi^2$ in the ordinary phase can also be extracted from the $\langle \phi\phi\rangle$ two-point function. The latter is computed to the $\epsilon^2$ order in \cite{Bissi:2018mcq} using analytical bootstrap and in \cite{Giombi:2020rmc} using equation of motion. Based on the two-point function, $a^{\rm ord}_{\phi^2}$ is expressed in terms of the OPE coefficient $C\equiv C_{\phi\phi\phi^2}$ as follows \cite{Bissi:2018mcq}
\begin{align}
C a^{\rm ord}_{\phi^2}  = -1+\frac{N+2}{2(N+8)} \epsilon + \frac{3(N+2)(3N+14)}{2(N+8)^3} \epsilon^2+\CO\left(\epsilon ^3\right)~.
\end{align}
Plugging in the OPE coefficient, c.f. eq.(3.17) of  \cite{Dey:2016mcs}, we find agreement with \eqref{aphi2}. This match provides a highly nontrivial check for the calculation of diagram $(f)$.

Because $a^{\rm ord}_{\phi^2} $ itself does not have the order $\epsilon$ term, it is better to do Pad\'e extrapolation for the product $C a^{\rm ord}_{\phi^2}$. Then we can plug in the numerical value of the OPE coefficient in $3d$ from conformal bootstrap \cite{Kos:2016ysd} to compute $a^{\rm ord}_{\phi^2}$ in the 3$d$ O$(N)$ model. In particular, in two dimensions, the one-point function of the $\varepsilon$ operator in the ordinary universality class of the Ising minimal model is known exactly $\langle \varepsilon (z)\rangle_{\rm ord} =-\frac{1}{2z}$ \cite{Cardy:1984bb, Kawai:2002pz} and the OPE coefficient is $C_{\sigma\sigma\varepsilon} = \frac{1}{2}$. This allows us to do two-sided Pade of $a^{\rm ord}_{\phi^2}$ in the $N=1$ case. We summarize our Pad\'e results for both $C a^{\rm ord}_{\phi^2}$ and $ a^{\rm ord}_{\phi^2}$ in Table \eqref{a2N}. They are in 
good quantitative agreement with the bootstrap calculation ($N=1,2,3$) \cite{Gliozzi:2015qsa} and the more recent fuzzy sphere calculation ($N=1$) \cite{Zhou:2024dbt}.

\begin{table}[t]
\centering
\renewcommand{\arraystretch}{1.2}
\begin{tabular}{|l|ccc|cc|cc|}
\hline
\multirow{2}{*}{}        & \multicolumn{3}{c|}{$N=1$}                                                               & \multicolumn{2}{c|}{$N=2$}                                  & \multicolumn{2}{c|}{$N=3$}                                  \\ \cline{2-8} 
                         & \multicolumn{1}{c|}{Pad\'e} & \multicolumn{1}{c|}{CB} & \multicolumn{1}{c|}{FS} & \multicolumn{1}{c|}{Pad\'e} & \multicolumn{1}{c|}{CB} & \multicolumn{1}{c|}{Pad\'e} & \multicolumn{1}{c|}{CB} \\ \hline
\multicolumn{1}{|c|}{$C a^{\rm ord}_{\phi^2} $ } & \multicolumn{1}{c|}{$-0.766$}   & \multicolumn{1}{c|}{$-0.789(3)$}         & $-$                         & \multicolumn{1}{c|}{$-0.735$}   & $-0.747(1)$                              & \multicolumn{1}{c|}{ $-0.710$}   &  $-0.710(1)$                             \\ \hline
\multicolumn{1}{|c|}{$a^{\rm ord}_{\phi^2} $ }  & \multicolumn{1}{c|}{$-0.728$}   & \multicolumn{1}{c|}{$-0.750(3)$}         &  $-0.74(4)$                        & \multicolumn{1}{c|}{$-1.07$}   & $-1.09(2)$                              & \multicolumn{1}{c|}{$-1.35$}   &  $-1.35(2)$                             \\ \hline
\end{tabular}
\caption{Pad\'e estimates of the one-point coefficient of $\phi^2$  in the ordinary boundary of the 3$d$ O$(N)$ model. The $N=2$ and $N=3$ data are obtained by the  one-sided [0,2] Pad\'e approximant because the [1,1] Pad\'e has a pole between $d=2$ and $d=3$. For the $N=1$ data, we average over all the available Pad\'e, i.e. the one-sided [0,2] Pad\'e and the two-sided [0,3], [1,2] and [2,1] Pad\'e. We also list  the results from conformal bootstrap (CB)  \cite{Gliozzi:2015qsa} and fuzzy sphere regularization (FS)  \cite{Zhou:2024dbt} as a reference. }
\label{a2N}
\end{table}

The $\epsilon$-expansion  for the $\phi^2$ one-point function is also available in the ``ordinary" (symmetry preserving) surface defect of the O$(N)$ model \cite{Diatlyk:2024ngd}. Converting the result in \cite{Diatlyk:2024ngd} to our convention gives for $N=1$:
\begin{align}
\text{Ordinary defect of Ising}: \quad \tilde a^{\rm ord}_{\phi^2}  = -0.942808 \epsilon + 0.130129 \epsilon ^2+0.174275 \epsilon ^3+\CO\left(\epsilon ^4\right)~.
\end{align}
In $3d$, the surface defect factorizes into a pair of ordinary boundary universality class. Therefore, applying Pad\'e resummation to $\tilde a^{\rm ord}_{\phi^2} $  and taking $\epsilon=1$ would yield an approximation of $ a^{\rm ord}_{\phi^2}  $. Averaging the three two-sided Pad\'e approximants, we obtain $\tilde a^{\rm ord}_{\phi^2}\approx -0.725 $ in 3$d$, which is also in good agreement with our boundary calculation.




\section{The symmetry breaking boundary condition}\label{extsec}

As reviewed in the introduction, the O$(N)$ model also admits symmetry breaking boundary conditions. Here we will focus on the so-called normal universality class, which corresponds to explicit symmetry breaking, and is defined by imposing boundary conditions such that $\phi^N$ has a non-zero one point function (this may be achieved by adding a boundary magnetic field in the $N$-th direction). Note that, since we work in $d=4-\epsilon$, the distinction between normal and extraordinary universality classes is not in fact very important, as the they are expected to be equivalent in $3<d<4$. In $d=3$, they are also expected to be equivalent for $N=1$, but not for $N>1$.

Due to the O$(N)$ symmetry breaking boundary condition, the field $\phi^N$ obtains a nonvanishing expectation value $\phi^N_{\rm vev}=\sqrt{\frac{d(d-2)}{4\lambda_0}}$ which can be obtained from the potential in the AdS action (\ref{Sor}). Expanding around this value, i.e. $\phi^N = \phi^N_{\rm vev}+\chi$,  yields the following action in the $d$-dimensional hyperbolic space $H^d$ \cite{ Giombi:2020rmc}
\begin{align}\label{Sex}
S &=-\frac{d^2(d-2)^2 }{64\lambda_0} {\rm Vol}(H^{d})+ \int d^{d} x\sqrt{g} \left[\frac{1}{2}\left(\partial\phi^a\right)^2+\frac{1}{2}\left(\partial\chi\right)^2+\frac{d(d-2)}{4}\chi^2\right.\nonumber\\
&+\left.\frac{\sqrt{d(d-2)\lambda_0}}{2}\left(\chi^3 +\phi^a\phi^a \chi\right)+\frac{\lambda_0}{4}\left((\phi^a\phi^a)^2+\chi^4+2\phi^a\phi^a\chi^2\right)\right]~,
\end{align}
where the index $a$ runs from $1$ to $N-1$. So there are $N-1$ massless fields with the boundary condition $\Delta_{\phi} =d-1$, and one massive scalar $\chi$ with the boundary scaling dimension being
\begin{align}
\Delta_{\chi} = \frac{d-1+\sqrt{3(d-1)^2-2}}{2}
\end{align}
which approaches 4 when $d=4$. In this section, we will compute the free energy of this theory to order $\epsilon$ at the WF fixed point and estimate the boundary central charge of the normal universality class in $d=3$ dimensions using Pad\'e resummations. We will also compute the leading order correction of the one-point function of $\phi^N$.

\subsection{The two-loop free energy}
First, the theory has a nonvanishing tree-level free energy, given by the first term of \eqref{Sex}
\begin{align}F_{\rm tree}^{\rm nor} = -\frac{d^2(d-2)^2 }{64\lambda_0} {\rm Vol}(H^{d})~.\end{align}
Next, the one-loop contribution is completely fixed by the mass of $\phi$ and $\chi$
\begin{align}F_{\rm 1-loop}^{\rm nor} = (N-1)F_1(\Delta_\phi) + F_1(\Delta_\chi)~.\end{align}
We can obtain the $\epsilon$-expansion of $F_1(\Delta_\phi)$ and $F_1(\Delta_\chi)$  numerically by using \eqref{F1mD} 
\begin{align}
&F_1(\Delta_\phi)=F_D-\frac{{\rm Vol}(H^{d})}{8\pi^2}\left(\frac{1}{\epsilon }+0.249205\, +0.430121 \epsilon \right)+\CO(\epsilon^2)~,\nonumber\\
&F_1(\Delta_\chi)=F_D-\frac{{\rm Vol}(H^{d})}{8\pi^2}\left(\frac{9}{\epsilon }-4.96867\, +5.34589 \epsilon \right)+\CO(\epsilon^2)~,
\end{align}
with $F_D$ given by \eqref{FDexp1}.

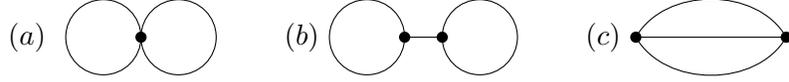
\begin{figure}
\centering
$(a)$\,\, \begin{tikzpicture} [baseline={(0,-0.1)}]
\draw[color=black] (0,0) circle [radius=0.5]; 
\draw[color=black] (1,0) circle [radius=0.5]; 
\draw (0.5,0) node[vertex]{} ;
\end{tikzpicture} 
\qquad
$(b)$\,\,\begin{tikzpicture} [baseline={(0,-0.1)}]
\draw[color=black] (0,0) circle [radius=0.5]; 
\draw[color=black] (1.5,0) circle [radius=0.5]; 
\draw (0.5,0) node[vertex]{} to (1,0) node[vertex]{};
\end{tikzpicture} 
\qquad
$(c)$\,\,\begin{tikzpicture} [baseline={(0,-0.1)}]
\draw (0,0) node[vertex]{} to (2,0) node[vertex]{} to [out=120, in=60] (0,0) to [out=-60, in=-120] (2,0);
\end{tikzpicture} 
\caption{The two-loop contributions to the free energy in the symmetry breaking phase.}
\label{SBdiagram}
\end{figure}

At the two-loop order, there are three types of  diagrams as shown in fig. \ref{SBdiagram}. 
The contact diagram $(a)$ in fig.\eqref{SBdiagram} comes from the quartic vertex of the $\phi^a$ fields
\begin{align}
\CF_2^{(a)} = \frac{\lambda_0 {\rm Vol}(H^{d})}{4}\left[(N^2-1)G_\phi(1)^2+3G_\chi(1)^2+2(N-1)G_\phi(1)G_\chi(1)\right],
\end{align}
where $G_{\phi/\chi}(1)$ denotes the coincident limit of the $\phi/\chi$ propagator, whose explicit expression is given by \eqref{GdXX} with $\Delta = \Delta_{\phi/\chi}$.
The diagram $(b)$ in fig.\eqref{SBdiagram} reads
\begin{align}\label{CF2def}
\CF_2^{(b)} &= -\frac{d(d-2)\lambda_0}{8} \left(3G_\chi(1)+(N-1) G_\phi(1)\right)^2\int_{X, Y}G_\chi(X,Y)~.
\end{align}
The  AdS integral in \eqref{CF2def} is a special case of $ \int_X G_\Delta (X, Y) $, which by symmetry is independent of $Y$. We denote it by  $A(\Delta)$ and we derive a differential equation of $A(\Delta)$ as follows. Consider $\int_{X, Y} G_\Delta (X, Y) G_\Delta (Y, Z)$. Integrating out $X$ first yields $A(\Delta)^2$. Instead, we can also integrate out $Y$ first. As we discussed in the previous section, the $Y$ integral gives a derivative of $G_\Delta(X, Z)$. More precisely, we have 
\begin{align}
A(\Delta)^2 = \int_X \left( \frac{\partial_\Delta}{d-1-2\Delta}G_\Delta(X, Z)\right) = \frac{A'(\Delta)}{d-1-2\Delta}~.
\end{align}
The solution of this first order differential equation is $A(\Delta) = \frac{1}{\Delta(\Delta+1-d)+\gamma}$ with $\gamma$ being a $\Delta$-independent integration constant. This constant can be fixed by evaluating $ \int_X G_\Delta (X, Y) $ for $\Delta=\frac{d}{2}$ with $G_\frac{d}{2}(X, Y)$ being an elementary function. One finds that the integration constant $\gamma$ appears to vanish and thus 
\begin{align}\label{tadpole}
\int_X G_\Delta(X, Y) = \frac{1}{\Delta(\Delta+1-d)}~.
\end{align}
Plugging \eqref{tadpole} into \eqref{CF2def} gives 
\begin{align}\label{CF2bfinal}
\CF_2^{(b)}=-\frac{\lambda_0 {\rm Vol}(H^{d})}{4}\left(3G_\chi(1)+(N-1) G_\phi(1)\right)^2 ~.
\end{align}
Finally, for the diagram $(c)$ in fig. \ref{SBdiagram}, we have
\begin{align}\label{F2C}
\CF_2^{(c)} &=- \frac{d(d-2)\lambda_0}{8}\int_{X, Y}\left[6 G_\chi(X, Y)^3+2(N-1)G_\phi(X, Y)^2G_\chi(X, Y)\right]\nonumber\\
&=-\frac{d(d-2)\lambda_0}{4}{\rm Vol}(H^{d}) {\rm Vol}(S^{d-1}) \int_1^\infty d\sigma  k_\epsilon(\sigma)~,\\
k_\epsilon(\sigma) &=  (\sigma^2-1)^{\frac{d}{2}-1}\left(3 G_\chi(\sigma)^2+(N-1)G_\phi(\sigma)^2\right) G_\chi(\sigma)~.
\end{align}
We will use the subtraction method to derive the $\epsilon$-expansion of $\int d\sigma k_\epsilon(\sigma)$. The method has been explained in the previous section.
In order to find the proper subtraction function $\hat k_\epsilon$, we need the behavior of  $G_\Delta(\sigma)$ near $\sigma=1$, which can be obtained using the following  formula of hypergeometric functions
\begin{align}\label{mF}
\mathbf{F} (a, b, c, z) = &\frac{\Gamma(c-a-b)}{\Gamma(c-a)\Gamma(c-b)}F(a,b,a+b-c+1,1-z)\nonumber\\
&+\frac{\Gamma(a+b-c)(1-z)^{c-a-b}}{\Gamma(a)\Gamma(b)}F(c-a,c-b,c-a-b+1,1-z)~.
\end{align}
In our case, we have $a=\frac{\Delta}{2}, b = \frac{\Delta+1}{2}, c= \Delta+\frac{3-d}{2}$ and $z= \frac{1}{\sigma^2}$. On the R.H.S, we can expand the hypergeometric functions as a series of $1-z= \frac{\sigma^2-1}{\sigma^2}$ and the most singular term comes from $(1-z)^{c-a-b}=(\frac{\sigma^2-1}{\sigma^2})^{\frac{2-d}{2}}$, which is roughly $\frac{1}{(\sigma-1)^{1-\epsilon/2}}$ around $\sigma=1$. For our purpose, we can simply replace the first hypergeometric function $F(a,b,a+b-c+1,1-z)$ by 1 and truncate the second hypergeometric function $F(c-a,c-b,c-a-b+1,1-z)$ to the linear order of $1-z$:
\small
\begin{align}\label{Gexp}
G_\Delta(\sigma) = \frac{\pi ^{\frac{\epsilon }{2}}}{(4\pi)^2\sigma^\Delta} \left(\frac{2^{\epsilon } \Gamma (\Delta ) \Gamma \left(\frac{\epsilon }{2}-1\right)}{\Gamma
   (\Delta +\epsilon -2)}- \Gamma \left(-\frac{\epsilon }{2}\right) \frac{(\sigma^2-1) (\Delta +\epsilon -2) (\Delta
   +\epsilon -1)+2 \sigma^2 \epsilon}{(\sigma^2-1)^{1-\frac{\epsilon }{2}} \sigma^{\epsilon }}\right)+\cdots
\end{align}
\normalsize
where ``$\cdots$'' denotes terms that vanish at $\sigma=1$ in the $\epsilon\to 0$ limit. Now we plug \eqref{Gexp} into $k_\epsilon$, and keep all the terms that do not vanish at $\sigma=1$ in the $\epsilon\to 0$ limit. This procedure gives one choice of  $\hat k_{\epsilon}$, 
which is a very complicated linear combination of $(\sigma^2-1)^{\#_1}\sigma^{\#_2}$. The integral of $\hat k_\epsilon$ can be carried out analytically for any $\epsilon$ using
\begin{align}
\int_1^\infty \frac{d\sigma}{(\sigma^2-1)^a\sigma^b} = \frac{\Gamma (1-a) \Gamma \left(a+\frac{b-1}{2}\right)}{2 \Gamma \left(\frac{1+b}{2}\right)}~.
\end{align}
The remaining integral is 
\begin{align}
d=4: \quad \int_1^\infty d\sigma (k_0 -\hat k_0) = \frac{5 \left(3 \pi ^2-38\right) N}{4608 \pi ^6}+\frac{40425 \pi ^2-401908}{2822400 \pi ^6}~.
\end{align}
Altogether, we get the two-loop part of the free energy
\begin{align}
F^{\rm nor}_{\rm 2-loop}=\CF_2^{(a)}+\CF_2^{(b)}+\CF_2^{(c)}=3\lambda {\rm Vol}(H^{d})\left[ \frac{3N+14}{128 \pi ^4 \epsilon } (1+\epsilon\xi)-\frac{15N+112}{(4 \pi) ^4}+\CO(\epsilon)\right]~,
\end{align}
where we have replaced the bare coupling $\lambda_0$ by the renormalized coupling $\lambda$ since our final result will be  valid up to the order of $\lambda$.
The $\epsilon^{-2}$ poles in $\CF_2^{(a)}, \CF_2^{(b)}$ and  $\CF_2^{(c)}$ cancel out in $F^{\rm nor}_{\rm 2-loop}$, and the remaining $\epsilon^{-1}$ pole in  $F^{\rm nor}_{\rm 2-loop}$ is removed by the tree level contribution $F_{\rm tree}^{\rm nor} = -\frac{ {\rm Vol}(H^{d})}{\lambda_0}$ because
\begin{align}
\frac{1}{\lambda_0} = \frac{1}{\lambda }-\frac{N+8}{8 \pi ^2 \epsilon }+\frac{3   (3 N+14)\lambda}{128 \pi ^4 \epsilon }+\CO(\lambda^2)~.
\end{align}
We evaluate the full two-loop free energy $F_{\rm nor} \equiv F_{\rm tree}^{\rm nor} + F_{\rm 1-loop}^{\rm nor} +F_{\rm 2-loop}^{\rm nor} $ at the WF fixed point \footnote{Compared to the ordinary boundary condition, we need to include the $\epsilon^3$ term of the fixed point $\lambda_\star$ because it will contribute an order $\epsilon$ term to the free energy via the factor $\frac{1}{\lambda_\star}$ in the tree level part.}
\small
\begin{align} \label{fixedlambda}
  \frac{\lambda_\star}{\pi^2}= &\,\frac{8\, \epsilon }{N+8}+ \frac{ 24 (3 N+14)}{(N+8)^3}\epsilon ^2-\frac{ 33N^3-110N^2-1760N-4544 + 96 \zeta(3) \left( N+8\right) \left( 5N+22\right)}{( N+8)^5} \epsilon^3 
\end{align}
\normalsize
and expand $F_{\rm nor}$ up to order $\epsilon$
\begin{align}\label{F2ext}
F_{\rm nor}&=-\frac{30(N+8)-N}{180\epsilon }-\frac{0.0057916N^2-2.6177 N-16.312}{N+8}\nonumber\\
&-\frac{\left(0.26513 N^4+10.658 N^3+162.62 N^2+1097.4 N+2591.9\right) \epsilon }{(N+8)^3}+\CO(\epsilon^2)~.
\end{align}
Note that the $1/\epsilon$ pole in this result is not related to a UV divergence (all divergences have cancelled), but is just due to the contribution of the classical action via its dependence on the coupling constant.  

A useful check of the result can be obtained by comparing to the large $N$ expansion. The free energy at the symmetry breaking phase at large $N$ is equal to the one-loop free energy of $N$ scalars of boundary scaling dimension $d-1$ \cite{Giombi:2020rmc}, which can be computed by using \eqref{F1mD}
\begin{align}
N F_1(d-1)  = N F_D+\frac{N{\rm Vol}(H^{d})}{2(4\pi)^{\frac{d}{2}}}\int^{d/2-1}_{0}\,d\hat\Delta\frac{(2\hat\Delta+1)\Gamma(\hat\Delta+\frac{d}{2})\Gamma(1-\frac{d}{2})}{\Gamma(2-\frac{d}{2}+\hat\Delta)}~.
\end{align}
After making the substitution $\hat\Delta\to (1-\epsilon/2)x$, the integral domain of $x$ becomes $[0, 1]$. Then we can expand the integral in $\epsilon$ and perform the $x$ integral (numerically) for each term. This procedure leads to $N F_1(d-1) = -N(\frac{29}{180\epsilon}+0.00579161+ 0.265129 \epsilon +\CO(\epsilon^2))$. This provides a  nontrivial check of the $\epsilon$-expansion result \eqref{F2ext}.

Combining the AdS free energy \eqref{F2ext} and the sphere free energy \eqref{Fsphere}, we get the  $\widetilde s$ function for the normal universality class of the O$(N)$ model \footnote{At this order, the curvature counter terms do not contribute.}
\begin{align}\label{tsext}
\widetilde s_{O(N)}^{\,\rm nor}(d) =& -\frac{N+8}{6\epsilon}+\frac{-0.012349 N^2+2.5652 N+16.312}{N+8}\nonumber\\
&-\frac{0.06614 N^4+4.2369 N^3+84.934 N^2+679.64 N+1749.7}{(N+8)^3}\epsilon~.
\end{align}

\subsubsection{Pad\'e approximation of the free energy}

When $N=1$, \eqref{tsext} reduces to $\widetilde s_{\rm Ising}^{\,\rm nor}(d)=-\frac{3}{2\epsilon }+2.0961\, -3.4548 \epsilon$.
On the other hand, in two dimensions, the normal universality class of the Ising model corresponds to fixing the boundary spin, either up or down.
$\widetilde s_{\rm Ising}^{\,\rm nor}$ thus approaches the logarithm of the corresponding boundary $g$ function of the Ising minimal model, which is known exactly \cite{Dorey:2009vg}
\begin{align}
\widetilde s_{\rm Ising}^{\,\rm nor}(2) = \log(g) =   -\frac{1}{2}\log 2~.
\end{align}
Imposing the $d=2$ value as a boundary condition allows us to use $[m,n]$ Pad\'e approximants for $\epsilon\widetilde s_{\rm Ising}^{\,\rm nor}(4-\epsilon)$ with total order $m+n = 3$. Averaging over the three two-sided Pad\'e approximants yields the boundary central charge of the normal (or extraordinary) universality class of the 3$d$ Ising model $c_{\rm Ising}^{\rm nor}  \approx - 1.43$,
which is just 0.7\% away from the fuzzy sphere result $ -1.44(6)$ \cite{Zhou:2024dbt}. In contrast, applying the one-sided [1,1] Pad\'e gives instead $c_{\rm Ising}^{\rm nor}\approx -1.35$, which deviates by about 6\% away from the fuzzy sphere calculation.

For $N>1$, only the one-sided [0,2] and [1,1] Pad\'e resummations are available. The former develops a pole in the physical region of $d$ as we increase $N$ while the latter is regular  for any $d>0$ and $N>1$. So we adopt the [1,1] Pad\'e, leading to $c_{O(2)}^{\rm nor}  \approx - 1.63$ for the O$(2)$ model and $c_{O(3)}^{\rm nor}  \approx - 1.92$ for the O$(3)$ model at the normal universality class. 

Let's also mention that the first order correction in the large $N$ expansion of the $c_{O(N)}^{\rm nor}$ was computed in \cite{Krishnan:2023cff}:  $c_{O(N)}^{{\rm nor},N} = -\frac{N}{2}-\frac{1}{16}+\CO(N^{-1})$. For the first  few values of $N$, the estimates are $c_{\rm Ising}^{{\rm nor},N} \approx -0.56,  c_{O(2)}^{{\rm nor},N} \approx -1.06$ and  $c_{O(3)}^{{\rm nor},N}\approx -1.56 $.

\subsection{The $\phi^N$ one-point function}

In this subsection,  we show how to compute the first-order correction to   $\langle \phi^N\rangle = \phi^N_{\rm vev} + \langle \chi\rangle$. At this order, we don't need to worry about the mixing with curvature terms and  
only the tadpole diagram \begin{tikzpicture} [baseline={(0,-0.1)}]
\draw[color=black] (0,0) circle [radius=0.2]; 
\draw (-0.5,0)  to (-0.2,0) node[vertex]{};
\end{tikzpicture} 
contributes to the $\chi$ one-point function, namely
\begin{align}
\langle\chi\rangle &= - \frac{\sqrt{d(d-2)\lambda_0}}{2}\left(3G_\chi(1)+(N-1)G_\phi(1)\right)\int_Y G_\chi(X, Y)~,\nonumber\\
&=\frac{\sqrt{\lambda_0}}{8 \sqrt{2} \pi ^2}\left(\frac{N+8}{ \epsilon }+\frac{N(4 \xi -5)}{8}+4 \xi -11+\CO\left(\epsilon\right)\right)
\end{align}
where we have used \eqref{tadpole} for the $Y$ integral. We don't need to include the wavefunction renormalization of $\chi$ as it starts at the quadratic order of $\lambda$.

Plugging in  $\phi^N_{\rm vev}=\sqrt{\frac{d(d-2)}{4\lambda_0}}$ and the bulk renormalization $\lambda_0 = \mu^\epsilon(\lambda + \frac{(N+8)\lambda^2}{8\pi^2\epsilon}+\cdots)$, we obtain the $\phi^N$ one-point function 
\begin{align}
\left\langle\phi^N \right\rangle= \frac{8-3 \epsilon }{4 \sqrt{2\lambda} } +\frac{ (N(2 \xi -1)+16 (\xi -2))}{32 \sqrt{2} \pi ^2}\sqrt{\lambda }~,
\end{align}
which at the WF fixed point reduces to 
\begin{align}
\left\langle\phi^N\right\rangle_{\rm f.p}=\sqrt{\frac{N+8}{  4\pi^2\epsilon }}+\frac{  \xi  (N+8)^2-2
   N (N+31)-308}{8 \pi \left(N+8\right)^{3/2} }\sqrt{\epsilon }+\CO\left(\epsilon ^{3/2}\right)
\end{align}
At the leading order of large $N$ it agrees with \cite{Giombi:2020rmc}, i.e. $\left\langle\phi^N\right\rangle_{\rm f.p}^2 = -N\Gamma(d-1)\Gamma(1-\frac{d}{2})/(4\pi)^{\frac{d}{2}}$.

To compare with the literature, we need to normalize the one-point function by the two-point function coefficient of $\phi^N$ without a boundary. More precisely, in $\mathbb R^d$, the $\phi^N$ two-point function takes the form \cite{Diatlyk:2024ngd}
\begin{align}
\langle \phi^N (x)\phi^N(0)\rangle = \frac{\CN_\phi^2}{x^{d-2+2\gamma_\phi}}, \quad \CN_\phi = \frac{1}{4\pi^2}+\frac{\epsilon \log(\pi e^{\gamma_E})}{8\pi^2}+\CO(\epsilon^2)~.
\end{align}
Altogether, in the BCFT  convention, the one-point coefficient of $\phi^N$ in the normal boundary condition should be 
\begin{align}
a_\phi^{\rm nor} = \frac{2^{\Delta_\phi}}{\CN_\phi}  \left\langle\phi^N\right\rangle_{\rm f.p}=2 \sqrt{\frac{N+8}{\epsilon}}-\frac{N^2+31 N+154 }{(N+8)^{3/2}}\sqrt{\epsilon}+\CO\left(\epsilon ^{3/2}\right)~.
\end{align}

 \begin{table}[t]
 \centering
\begin{tabular}{|l|cll|cl|cl|}
\hline
\multirow{2}{*}{}       & \multicolumn{3}{c|}{$N=1$}                                                           & \multicolumn{2}{c|}{$N=2$}                          & \multicolumn{2}{c|}{$N=3$}                          \\ \cline{2-8} 
                        & \multicolumn{1}{c|}{Pad\'e} & \multicolumn{1}{c|}{MC} &  \multicolumn{1}{c|}{FS}                & \multicolumn{1}{c|}{Pad\'e} & \multicolumn{1}{c|}{MC}           & \multicolumn{1}{c|}{Pad\'e} &\multicolumn{1}{c|}{MC}           \\ \hline
\multicolumn{1}{|c|}{$a_\phi^{\rm nor}$ } & \multicolumn{1}{c|}{2.71}   & \multicolumn{1}{c|}{2.60(5)}         & \multicolumn{1}{c|}{2.58(16)} & \multicolumn{1}{c|}{3.01169}   & \multicolumn{1}{c|}{2.880(2)} & \multicolumn{1}{c|}{3.22339}   & \multicolumn{1}{c|}{3.136(2)} \\ \hline
\end{tabular}
\caption{Pad\'e estimate of the one-point coefficient of $\phi^N$ in the normal boundary condition of the 3$d$ O$(N)$ model. The $N=2$ and $N=3$ data are obtained by the  one-sided [0,1] Pad\'e approximant. For the $N=1$ data, we average over the two-sided [1,1] and [0,2] Pad\'e. We also list  the results from Monte-Carlo simulation (MC)   \cite{Toldin:2021kun} and fuzzy sphere regularization (FS)  \cite{Zhou:2024dbt} as a reference. }
\label{aphiN}
\end{table}

 For $N=1$, the  one-point function of the $\sigma$ operator with the symmetry breaking (normal) boundary condition of the Ising minimal model is known exactly $\langle \sigma (z)\rangle_{\rm nor} =(2/z)^{\frac{1}{8}}$ \cite{Cardy:1984bb, Kawai:2002pz}.  
 So we can apply a two-sided Pad\'e extrapolation to $\sqrt{\epsilon}a_\phi^{\rm nor}$.  For higher $N$, only the one-sided [0,1] Pad\'e is available. We summarize the results in Table \ref{aphiN}. They are in good  agreement with the Monte-Carlo simulation ($N = 1,2,3$)\cite{Toldin:2021kun}  and the
fuzzy sphere calculation ($N = 1$) \cite{Zhou:2024dbt}. We expect that including further sub-leading order corrections may lead to a better match with the available numerical methods.


\section*{Acknowledgements}
The work of S.G. and Z.S. is supported in part by the US National Science Foundation Grant No. PHY-2209997. Z.S. is also supported in part by the Simons Foundation Grant No. 917464.

\appendix

\section{A direct calculation of $F_D$}\label{FDexp}
In this appendix, we compute the free energy $F_D$ of a conformally coupled scalar with Dirichlet boundary condition in AdS$_{d}$ by using the spectral integral \eqref{F1def} 
\begin{align}\label{FDint}
F_D =\frac{{\rm Vol}(H^{d})}{2(4\pi)^{\frac{d}{2}} \Gamma(\frac{d}{2})}\int_\mathbb{R}d\lambda\CP_\epsilon(\lambda)  \log\left(\lambda^2+\frac{1}{4}\right), \quad \CP_\epsilon(\lambda) \equiv \frac{\Gamma(\frac{d-1}{2}+ i\lambda)}{\Gamma( i\lambda)}\frac{\Gamma(\frac{d-1}{2}- i\lambda)}{\Gamma( -i\lambda)}
\end{align}
We will obtain its $\epsilon$ expansion up to order $\epsilon$. This method can be easily generalized to higher orders and other masses.

The strategy is the same as computing $\CF_2^{(c)}$, cf. \eqref{F2C}. We need to find some function $\hat\CP_\epsilon(\epsilon)$ such that both 
\begin{align}
\int_\mathbb{R}d\lambda\left(\CP_\epsilon(\lambda)-\hat\CP_\epsilon(\lambda)\right)  \log\left(\lambda^2+\frac{1}{4}\right), \quad \int_\mathbb{R}d\lambda\partial_\epsilon\left(\CP_\epsilon(\lambda)-\hat\CP_\epsilon(\lambda)\right)  \log\left(\lambda^2+\frac{1}{4}\right)
\end{align}
are convergent for $\epsilon\in[0, \epsilon_c)$ (the explicit value of $\epsilon_c$ is not important), and the integral of $\hat\CP_\epsilon(\epsilon)\log\left(\lambda^2+\frac{1}{4}\right)$ can be evaluated analytically. For a given $\hat\CP_\epsilon$, we can determine the integral of $\CP_\epsilon(\epsilon)\log\left(\lambda^2+\frac{1}{4}\right)$  up to order $\epsilon$ as follows
\begin{align}\label{PQ}
\int_\mathbb{R}d\lambda\CP_\epsilon(\lambda)  \log\left(\lambda^2+\frac{1}{4}\right)&=\int_\mathbb{R}d\lambda\left[\left(\CP_0(\lambda)  -\hat\CP_0(\lambda)\right)+\epsilon \left(\CP'_0(\lambda)  -\hat\CP'_0(\lambda)\right)\right]\log\left(\lambda^2+\frac{1}{4}\right)\nonumber\\
&+ \int_\mathbb{R}d\lambda\hat\CP_\epsilon(\lambda)\log\left(\lambda^2+\frac{1}{4}\right)+\CO(\epsilon^2)~,
\end{align}
where the prime denote the derivative with respect to $\epsilon$.

In order to find such a $\hat\CP_\epsilon(\lambda)$, we first analyze the large $\lambda$ behavior of $\CP_\epsilon$ because $\lambda=\pm\infty$ are the only singularity of the original integral \eqref{FDint}. Using the Stirling's formula for $\Gamma$ functions, we get 
\begin{equation}
\begin{split}\label{CPasym}
\CP_\epsilon(\lambda)&=|\lambda|^{d-1}\bigg[1+\frac{(d-3) (d-2) d}{24 \lambda^2}+\frac{(d-5)(d-4) (d-3) (d-2) (d-1)  (5 d-3)}{5760 \lambda^4}\\
&\quad+\frac{(d-7)(d-6) (d-5) (d-4) (d-3) (d-2) (d-1)  \left(35 d^2-28 d+9\right)}{2903040
   \lambda^6}+\CO\left(\frac{1}{\lambda^8}\right)\bigg]~.
 \end{split}
\end{equation}
The function $\hat\CP_\epsilon(\lambda)$ should have the same large $\lambda$ asymptotic behavior up to $\lambda^{d-7}$. A convenient choice of $\hat\CP_\epsilon(\lambda)$  is thus 
\begin{align}\label{CQexp}
\hat\CP_\epsilon(\lambda) &=\left(\lambda^2+\frac{1}{4}\right)^{\frac{d-1}{2}}+ \frac{(d-1) \left(d^2-5 d+3\right)}{24\left(\lambda^2+\frac{1}{4}\right)^{\frac{3-d}{2}}}\nonumber\\
 &+\frac{(d-3) (d-1) \left(5 d^4-58 d^3+193 d^2-164 d-15\right) }{5760 \left(\lambda^2+\frac{1}{4}\right)^{\frac{5-d}{2}}}\\
   &+\frac{(d-5) (d-3) (d-1) \left(35 d^6-693 d^5+4706 d^4-12561 d^3+9776 d^2+2445 d+189\right)}{2903040\left(\lambda^2+\frac{1}{4}\right)^{\frac{7-d}{2}}}~.\nonumber
\end{align}
Here we choose $\hat\CP_\epsilon(\lambda)$ to be a function of $\lambda^2+\frac{1}{4}$ because the integral of $(\lambda^2+\frac{1}{4})^a \log(\lambda^2+\frac{1}{4})$ can be evaluated analytically
\begin{align}
\int_{\mathbb R} d\lambda\left(\lambda^2+\frac{1}{4}\right)^a\log\left(\lambda^2+\frac{1}{4}\right)= -\frac{\sqrt{\pi }\, \Gamma \left(-a-\frac{1}{2}\right) \left(\psi^{(0)}\left(-a-\frac{1}{2}\right)-\psi^{(0)}(-a)+\log (4)\right)}{2^{2a+ 1} \Gamma (-a)}~,
\end{align}
where $\psi^{(0)}$ is the digamma function.
Plugging \eqref{CQexp} into \eqref{PQ} and performing the first integral numerically, we get 
\begin{align}
F_D =\frac{1}{180 \epsilon }-0.0010552-0.0031491 \epsilon +\mathcal O\left(\epsilon ^2\right)~.
\end{align}
It agrees with \eqref{FDexp1} to the order of $\epsilon$. To obtain the higher order terms in $F_D$, it suffices to extend $\hat\CP_\epsilon$ such that it matches the asymptotic behavior of $\CP_\epsilon$ to higher orders. For any $F_1(\Delta)$, one  needs to expand $\hat\CP_\epsilon$ in $\lambda^2+(\Delta-\frac{d-1}{2})^2$ and the remaining steps are exactly the same.

\section{Harmonic expansion of CFT 2-point function on the cylinder}\label{harm}
Consider a CFT 2-point function on the cylinder $\mathbb R\times S^{d-1}$
\begin{align}
\CG^\Delta_{\rm cyl}(t, \omega_1; 0, \omega_2) &= \bigl( \cosh t - \omega_1 \cdot \omega_2 \bigr)^{-\Delta} =\frac{1}{\Gamma(\Delta)}  \int_0^\infty \frac{ds}{s} s^\Delta e^{-s(\cosh t - \omega_1 \cdot \omega_2)}  \, ,
\end{align} 
where $t$ is the coordinate of $\mathbb R$ and $\omega$ denotes a point on $S^{d-1}$.
 We will  write it in terms of Fourier modes in the $t$ direction and spherical harmonics in the angular directions. Let us first compute the Fourier transform of the $t$-dependent factor in the integrand:
\begin{align} \label{FTexpcosh}
 \int_{\mathbb R} dt \, e^{-i t\nu} \, e^{-s \cosh t}=
 \int_{\mathbb R_+} \frac{dr}{r} r^{-i\nu} \, e^{-\frac{s}{2}(r+r^{-1})}  = 2 K_{i \nu}(s) \, .
\end{align}
Next we expand the angular part in spherical harmonics:
\begin{align}
 e^{s \, \omega_1 \cdot \omega_2} = 2 \pi^\frac{d}{2} \sum_{\ell m} \left(\tfrac{s}{2}\right)^{-\alpha} I_{\alpha+\ell}(s) \, Y_{\ell m}(\omega_1) \, Y_{\ell m}^*(\omega_2) \, , 
\end{align}
where $\alpha=\frac{d}{2}-1$ and $I_{\alpha+\ell}(s)$ is a Bessel $I$ function. 
Therefore the 2-point function can be expanded as follows:
\begin{align}
\CG^\Delta_{\rm cyl}(t, \omega_1; 0, \omega_2) = \sum_{\ell m} \int_{\mathbb R} \frac{d\nu}{2\pi} \, \CK_\Delta(\nu,\ell) \, e^{i\nu t}\, Y_{\ell m}(\omega_1) Y^*_{\ell m}(\omega_2)~,
\end{align}
with the kernel function $\CK_\Delta(\nu,\ell)$ being
\begin{align}
 \CK_\Delta(\nu,\ell) &= \frac{  2^{2+\alpha} \, \pi^{\frac{d}{2}} }{\Gamma(\Delta)}  \int_0^\infty \frac{ds}{s} \, s^{\Delta-\alpha} \, K_{i \nu}(s) \, I_{\alpha+\ell}(s) \, \\
 &= \frac{2^\Delta \pi^{\frac{d}{2}} \Gamma(\frac{d}{2}-\Delta)}{\Gamma(\Delta)}
  \frac{\Gamma\bigl( \frac{1}{2}(\Delta+\ell-i \nu) \bigr) \, \Gamma\bigl( \frac{1}{2}(\Delta+\ell+i \nu) \bigr)}{\Gamma\bigl( \frac{1}{2}(\bar\Delta+\ell+i \nu) \bigr) \, \Gamma\bigl( \frac{1}{2}(\bar\Delta+\ell-i \nu) \bigr)} \, ,
\end{align} 
where $\bar\Delta \equiv d-\Delta$. Integrating over $\omega_1$ or $\omega_2$ projects the sum onto $\ell=0$:
\begin{align}
\int d^{d-1}\omega_1 d^{d-1}\omega_2\, \CG^\Delta_{\rm cyl}(t, \omega_1; 0, \omega_2) = {\rm Vol}(S^{d-1})\int_{\mathbb R} \frac{d\nu}{2\pi} \, \CK_\Delta(\nu,0) \, e^{i\nu t}~.
\end{align}
In particular, for $\Delta=d-2$ we have 
\begin{align}\label{FD}
&\int_{S^{d-1}}\frac{ d^{d-1}\omega_1 d^{d-1}\omega_2}{(\cosh t-\omega_1\cdot \omega_2)^{d-2}}=\int_{\mathbb R}d\nu\, \CK(\nu) e^{i\nu t}~,\nonumber\\
&  \CK(\nu) =\frac{2^{d-2} \pi ^{d-1} \Gamma \left(2-\frac{d}{2}\right)}{\Gamma (d-2) \Gamma \left(\frac{d}{2}\right)}\frac{\Gamma(\frac{d-2}{2}+i\frac{\nu}{2})\Gamma(\frac{d-2}{2}-i\frac{\nu}{2})}{\Gamma(1+i\frac{\nu}{2})\Gamma(1-i\frac{\nu}{2})}~.
\end{align}
The function $\CK(\nu) $ has a pole at $d=4$. The $\nu$ integral is convergent for $d<3$.

Another interesting example is $\Delta=\alpha$. In this case, $\CK_\alpha(\nu)$ is proportional to $\frac{1}{\alpha^2+\nu^2}$ and hence we can evaluate the $\nu$ integral analytically
\begin{align}\label{FDexact}
&\int_{S^{d-1}}\frac{ d^{d-1}\omega_1 d^{d-1}\omega_2}{(\cosh t-\omega_1\cdot \omega_2)^{\frac{d-2}{2}}}=\frac{2^{\frac{d}{2}+1} \pi ^d}{\Gamma \left(\frac{d}{2}\right)^2} e^{- \alpha |t|}~.
\end{align}

\section{An integral representation of sphere propagators}
Let's consider a heavy scalar field of mass $M$ on a unit sphere $S^{d-1}$, where ``heavy'' means $M^2=\frac{(d-2)^2}{4}+\lambda^2>\frac{(d-2)^2}{4}$. Then the Green function associated to this field is 
\begin{align}
\mathbb G_\lambda(\omega_1,\omega_2)=\sum_{\ell,m} \frac{Y_{\ell m}(\omega_1)\, Y_{\ell m}(\omega_2)^*}{\ell(\ell+d-2)+M^2}~.
\end{align}
Performing the sum over $m$ yields a Gegenbauer polynomial in terms of $\omega_1\cdot\omega_2$
\begin{align}\label{G1}
\mathbb G_\lambda(\omega_1,\omega_2)&=\frac{\Gamma(\alpha)}{4\pi^{\alpha+1}}\,\sum_\ell\frac{2(\ell+\alpha)}{(\ell+\alpha)^2+\lambda^2}\, C^{\alpha}_\ell(\omega_1\cdot\omega_2)\nonumber\\
&=\frac{\Gamma(\alpha)}{4\pi^{\alpha+1}}\,\sum_\ell\,\left(\frac{1}{\ell+\alpha+i\lambda}+\frac{1}{\ell+\alpha-i\lambda}\right)\, C^{\alpha}_\ell(\omega_1\cdot\omega_2)~.
\end{align}
To proceed, we need the generating function of Gegenbauer polynomials
\begin{align}\label{gen1}
\left(1-2\,\omega_1\cdot\omega_2 \, z+z^2\right)^{-\alpha}=\sum_{\ell}\,  C^{\alpha}_\ell(\omega_1\cdot\omega_2) \, z^\ell~.
\end{align}
One useful observation is that, if we have an operation that maps the R.H.S of \eqref{gen1} to the R.H.S of \eqref{G1}, then the Green function $\mathbb G_\lambda(\omega_1,\omega_2)$ can be related to $\left(1-2\,\omega_1\cdot\omega_2 \, z+z^2\right)^{-\alpha}$ by the same operation. It is easy to construct such an operation 
\begin{align}
\mathbb F\{g\}\equiv \int_0^1 \frac{dz}{z}\, \left(z^{\alpha+i\lambda}+z^{\alpha-i\lambda}\right) \, g(z)~.
\end{align}
whose action on $z^\ell$ is $
z^\ell\to \frac{1}{\ell+\alpha+i\lambda}+\frac{1}{\ell+\alpha-i\lambda}$.
$\mathbb F$ leads to the following integral representation of the sphere Green function 
\begin{align}\label{tint}
\mathbb G_\lambda(\omega_1,\omega_2)&=\frac{\Gamma(\alpha)}{4\pi^{\alpha+1}}\, \int_0^1 \frac{dz}{z}\, \frac{z^{\alpha+i\lambda}+z^{\alpha-i\lambda}}{\left(1-2\,\omega_1\cdot\omega_2 z+z^2\right)^{\alpha}}=\frac{\Gamma(\alpha)}{(2\pi)^{\alpha+1}}\, \int_0^\infty\, dt \, \frac{\cos(\lambda t)}{(\cosh t-\omega_1\cdot\omega_2)^{\alpha}}~.
\end{align}
Evaluating the $t$ integral in the coincident limit yields
\begin{align}\label{tint2}
\mathbb G_\lambda(\omega,\omega) = \frac{1}{(2\pi)^{\alpha+1}}\int_0^\infty \frac{ds}{s}s^\alpha e^s K_{i\lambda}(s)
=\frac{\Gamma(\frac{3-d}{2})}{(4\pi)^{\frac{d-1}{2}}} \frac{\Gamma(\frac{d-2}{2}+i\lambda)\Gamma(\frac{d-2}{2}-i\lambda)}{\Gamma(\frac{1}{2}+i\lambda)\Gamma(\frac{1}{2}-i\lambda)}~.
\end{align}
where we have used \eqref{FTexpcosh}.

As an application, let's consider the sphere integral of $\mathbb  G_{\lambda_1}(\omega_1,\omega_2) \mathbb  G_{\lambda_2}(\omega_1,\omega_2)$ over $\omega_2\in S^{d-1}$. On the one hand, we have 
\begin{align}\label{Oh}
\int d^{d-1}\omega_2 &\mathbb  G_{\lambda_1}(\omega_1,\omega_2)\mathbb   G_{\lambda_2}(\omega_1,\omega_2) = \int d^{d-1}\omega_2 \langle \omega_1|\frac{1}{-\nabla^2+M_1^2}|\omega_2\rangle \langle \omega_2 |\frac{1}{-\nabla^2+M_2^2}|\omega_1\rangle \nonumber\\
&=\langle \omega_1|\frac{1}{(-\nabla^2+M_1^2)(-\nabla^2+M_2^2)}|\omega_1\rangle = \frac{\mathbb G_{\lambda_1}(\omega_1, \omega_1)-\mathbb G_{\lambda_2}(\omega_1, \omega_1)}{\lambda_2^2-\lambda_1^2}~,
\end{align}
where we define $M_i^2\equiv \alpha^2+\lambda_i^2$ in the intermediate steps. On the other hand, plugging \eqref{tint} into this integral gives 
\small
\begin{align}\label{OOh}
\int d^{d-1}\omega_2 &\mathbb  G_{\lambda_1}(\omega_1,\omega_2)\mathbb   G_{\lambda_2}(\omega_1,\omega_2) =\frac{\Gamma(\alpha)^2}{(2\pi)^{d}}\int_{0}^\infty dt_1 dt_2 \int d^{d-1}\omega_2 \frac{\cos(\lambda_1 t_1)}{(\cosh t_1-\omega_1\cdot\omega_2)^{\alpha}} \frac{\cos(\lambda_2 t_2)}{(\cosh t_2-\omega_1\cdot\omega_2)^{\alpha}}~.
\end{align}
Combining \eqref{tint2}, \eqref{Oh} and \eqref{OOh}, we obtain the following double Fourier transformation 
\begin{align}\label{tK}
\int\frac{d^{d-1}\omega_2}{(\cosh t_1-\omega_1\cdot\omega_2)^{\alpha}(\cosh t_2-\omega_1\cdot\omega_2)^{\alpha}} = \int_{\mathbb R} d\lambda_1 d\lambda_2\, e^{i\lambda_1 t_1+i\lambda_2 t_2} \widetilde\CK(\lambda_1, \lambda_2) ~,
\end{align}
where 
\begin{align}\label{tKdef}
\widetilde\CK(\lambda_1, \lambda_2) = \frac{(4\pi)^\alpha }{\Gamma (\alpha) \Gamma
   (2\alpha)\cos(\frac{d\pi}{2})} \frac{\frac{\Gamma(\alpha+i\lambda_1)\Gamma(\alpha-i\lambda_1)}{\Gamma(\frac{1}{2}+i\lambda_1)\Gamma(\frac{1}{2}-i\lambda_1)}-\frac{\Gamma(\alpha+i\lambda_2)\Gamma(\alpha-i\lambda_2)}{\Gamma(\frac{1}{2}+i\lambda_2)\Gamma(\frac{1}{2}-i\lambda_2)}}{\lambda^2_1-\lambda_2^2}~.
\end{align}
Depending on the sign of $t_1$ and $t_2$, we can close  the $\lambda_1$ and $\lambda_2$ contour in the upper or lower half plane, which yields
\begin{align}\label{ddtK}
\int\frac{d^{d-1}\omega_1d^{d-1}\omega_2}{(\cosh t_1-\omega_1\cdot\omega_2)^{\alpha}(\cosh t_2-\omega_1\cdot\omega_2)^{\alpha}} = \frac{(2\pi)^d}{\Gamma(\frac{d}{2})^2} \,
 \frac{ F\left(\alpha,2\alpha;\alpha+1, e^{-|t_1|-|t_2|}\right)}{e^{\alpha (|t_1|+|t_2|)}}~.
\end{align}
In $d=4$ dimensions, it reduces to 
\begin{align}
\int\frac{d^{3}\omega_1d^{3}\omega_2}{(\cosh t_1-\omega_1\cdot\omega_2)(\cosh t_2-\omega_1\cdot\omega_2)} = \frac{(2\pi)^4}{e^{|t_1|+|t_2|}-1}~.
\end{align}

\section{$\langle\phi^2\rangle$ on the round sphere}\label{sphereagain}
In this appendix, we put the O($N$) model on $S^d$ and compute the one-point function of the bare $\phi_0^2$ operator. Following \cite{Fei:2015oha}, we work in the stereograhic coordinates of $S^d$, i.e. $ds^2 = \mathbb{\Omega}(x)^2 dx^2$, where $x^i\in\mathbb R^d$ and $\mathbb{\Omega}(x) = \frac{2}{1+x^2}$.
The $\phi$ propagator in the stereographic coordinates reads
\begin{align}
\left\langle \phi_0^I(x)\phi_0^J (y)\right\rangle=\frac{\delta^{IJ}\CN_d}{\left( \mathbb{\Omega}(x)\mathbb{\Omega}(y)(x-y)^2\right)^\frac{d-2}{2}}, \quad \CN_d = \frac{ \Gamma \left(\frac{d}{2}-1\right)}{4 \pi ^{d/2}}~.
\end{align}

In dimensional regularization, the leading contribution to $\left\langle \phi_0^2 (0)\right\rangle$ appears at the $\lambda^2$ order
\begin{align}
\left\langle \phi_0^2\right\rangle = 
\begin{tikzpicture}[baseline={(0,-0.1)}]
\draw (0,0) node[vertex]{} to (2,1)   node[vertex]{}  to  (2,-1) node[vertex]{} to (0,0)  ;
\draw (2,1) to [out =-60, in =60] (2,-1);
\draw (2,1) to [out =-120, in =120] (2,-1);
\end{tikzpicture}  = 2 N (N + 2) \lambda_0^2  \CN_d ^5 \int \frac{\mathbb \Omega(x_1)^d \mathbb \Omega(x_2)^d d^d x_1 d^d x_2}{(\mathbb \Omega(0)\mathbb \Omega(x_1)x_1^2)^{\frac{d-2}{2}}(\mathbb \Omega(0)\mathbb \Omega(x_2)x_2^2)^{\frac{d-2}{2}}(\mathbb \Omega(x_1)\mathbb \Omega(x_2)x_{12}^2)^{\frac{3(d-2)}{2}}}~.
\end{align}
A useful trick for dealing with such sphere integrals is the inversion $x^i \to \frac{x^i}{x^2}$ \cite{Cardy:1988cwa}. After performing this transformation, we get 
\begin{align}
\left\langle \phi_0^2\right\rangle =2^{-1+3\epsilon} N (N + 2) \lambda_0^2  \CN_d ^5\int\frac{ d^d x_1 d^d x_2}{(1+x_1^2)^\epsilon (1+x_2^2)^\epsilon x_{12}^{3(d-2)}}~,
\end{align} 
which is exactly solvable (see for example appendix B of \cite{Fei:2015oha})
\begin{align}\label{phi2S}
\left\langle \phi_0^2\right\rangle =  - N (N + 2) \lambda_0^2\frac{ \Gamma \left(1-\frac{\epsilon }{2}\right)^4 \Gamma \left(\frac{3 \epsilon
   }{2}-1\right)}{2^{10-3 \epsilon }  \pi ^{6-\frac{3 \epsilon }{2}}(\epsilon-1)(\epsilon-2) \Gamma (\epsilon )}~.
\end{align}
The R.H.S of \eqref{phi2S} is finite in the $\epsilon\to 0$ limit. So at this order, the one-point function of the renormalized operator $[\phi^2]_R =Z_{\phi^2}^{-1}\phi_0^2 $ is simply
\begin{align}
\left\langle [\phi^2]_R \right\rangle =  - N (N + 2) \lambda^2\frac{ \Gamma \left(1-\frac{\epsilon }{2}\right)^4 \Gamma \left(\frac{3 \epsilon
   }{2}-1\right)}{2^{10-3 \epsilon }  \pi ^{6-\frac{3 \epsilon }{2}}(\epsilon-1)(\epsilon-2) \Gamma (\epsilon )} \stackrel{\lambda = \lambda_\star}{=}\frac{N(N+2)\epsilon^2}{48\pi^2(N+8)^2}+\CO(\epsilon^3)~.
\end{align}

By conformal symmetry, a scalar primary operator in a CFT should have a vanishing one-point function on sphere. The calculation above shows that, unlike in flat space, $[\phi^2]_R$ is {\it not} a primary operator on sphere. In order to properly define a primary operator on sphere, we have to add a finite curvature counterterm to  $[\phi^2]_R$ 
\begin{align}
[\phi^2]_{S^d} \equiv [\phi^2]_R +\kappa \mu^{-\epsilon}\, {\cal R}_{S^d}, \quad {\cal R}_{S^d} = d(d-1)~,
\end{align}
where $\mu$ is an arbitrary mass scale and $\kappa$ is a constant fixed by the condition $\langle [\phi^2]_{S^d}\rangle =0$. At the $\epsilon^2$ order, we find $\kappa$ to be
\begin{align}\label{defb}
\kappa= -\frac{N(N+2)\epsilon^2}{576\pi^2(N+8)^2}+\CO(\epsilon^3)~.
\end{align}

On the hyperbolic space $H^d$,  the analogue of $ [\phi^2]_{S^d} $ should be 
 \begin{align}\label{phi2Hd}
[\phi^2]_{H^d} \equiv [\phi^2]_R +\kappa \mu^{-\epsilon}\, {\cal R}_{H^d}, \quad {\cal R}_{H^d} = -d(d-1)~,
\end{align}
with the coefficient $\kappa$ given by \eqref{defb}.

\bibliography{refs}
\bibliographystyle{utphys}

\end{document}